\documentclass{IEEEtran}
\IEEEoverridecommandlockouts
\usepackage{acronym}
\acrodef{6dma}[6DMA]{six-dimensional movable antenna}
\acrodef{em}[EM]{electromagnetic}
\acrodef{6g}[6G]{six-generation}
\acrodef{ap}[AP]{access point}
\acrodef{mimo}[MIMO]{multi-input multi-output}
\acrodef{siso}[SISO]{single-input single-output}
\acrodef{cpu}[CPU]{central processing unit}
\acrodef{lpu}[LPU]{local processing unit}
\acrodef{lcs}[LCS]{local Cartesian coordinate system}
\acrodef{gcs}[GCS]{global Cartesian coordinate system}
\acrodef{3d}[3D]{three-dimensional}
\acrodef{csi}[CSI]{channel state information}
\acrodef{bs}[BS]{base station}
\acrodef{mmse}[MMSE]{minimum mean-square error}
\acrodef{mse}[MSE]{mean-square error}
\acrodef{uatf}[UatF]{use-and-then-forget}
\acrodef{FP}[FP]{fractional programming}
\acrodef{cssca}[CSSCA]{constrainted stochastic successive convex approximation}
\acrodef{sinr}[SINR]{signal-to-interference-plus-noise ratio}
\acrodef{ao}[AO]{alternating optimization}
\acrodef{fpa}[FPA]{fixed-position antenna}
\acrodef{rf}[RF]{radio frequency}
\acrodef{fpa}[FPA]{fixed-position antenna}
\acrodef{dof}[DoFs]{degrees of freedom}
\acrodef{los}[LoS]{line of sight}
\acrodef{mumimo}[MU-MIMO]{multi-user MIMO}
\acrodef{ma}[MA]{movable antenna}
\acrodef{lmmse}[LMMSE]{local minimum mean-square error}
\acrodef{uav}[UAV]{unmanned aerial vehicle}
\acrodef{fas}[FAS]{fluid antenna system}
\acrodef{saa}[SAA]{sampled average approximation}
\acrodef{cdf}[CDF]{cumulative distribution function} 
\usepackage{amsmath}
\usepackage{graphicx} 
\usepackage{epstopdf}
\usepackage{amssymb}
\usepackage{amsfonts}
\usepackage{amsthm}
\usepackage{cite}
\usepackage{bm,comment}
\usepackage{algorithm}
\usepackage{algorithmic}

\usepackage{subeqnarray}
\usepackage{subfigure}
\usepackage{multicol}
\usepackage{multirow}
\usepackage{diagbox}
\usepackage{slashbox}
\usepackage{stfloats}
\usepackage{float}
\usepackage{color}
\usepackage{cases}
\usepackage{CJK}
\usepackage{bm}
\usepackage{bbm}
\usepackage{placeins}

\usepackage[colorlinks,linkcolor=blue,citecolor=red]{hyperref}
\usepackage{lipsum}

\newtheorem{remark}{\bf{Remark}}

\usepackage{geometry}
\def\BibTeX{{\rm B\kern-.05em{\sc i\kern-.025em b}\kern-.08em
    T\kern-.1667em\lower.7ex\hbox{E}\kern-.125emX}}

\begin{document}
\title{6D Movable Antenna Enhanced Cell-free MIMO: Two-timescale Decentralized Beamforming and Antenna Movement Optimization}

\author{Yichi Zhang, Yuchen Zhang, \textit{Member}, \textit{IEEE}, Wenyan Ma, \textit{Graduate Student Member}, \textit{IEEE}, Lipeng Zhu, \textit{Member}, \textit{IEEE}, Jianquan Wang, \textit{Member}, \textit{IEEE}, Wanbin Tang, \textit{Member}, \textit{IEEE}, and Rui Zhang, \textit{Fellow}, \textit{IEEE}

\thanks{

Yichi Zhang and Wanbin Tang are with the National Key Laboratory of Wireless Communications, University of Electronic Science and Technology of China, Chengdu 611731, China (e-mail: yczhang@std.uestc.edu.cn, wbtang@uestc.edu.cn).

Yuchen Zhang is with the Electrical and Computer Engineering Program, Division of Computer, Electrical and Mathematical Sciences and Engineering (CEMSE), King Abdullah University of Science and Technology (KAUST), Thuwal, 23955-6900, Kingdom of Saudi Arabia (e-mail: yuchen.zhang@kaust.edu.sa).

Lipeng Zhu is with the School of Interdisciplinary Science, Beijing Institute of Technology, Beijing 100081, China (e-mail: lipzhu@outlook.com).

Wenyan Ma, and Rui Zhang are with the Department of Electrical and Computer Engineering,
National University of Singapore, Singapore 117583 (e-mail:
wenyan@u.nus.edu; elezhang@nus.edu.sg).

Jianquan Wang is with the National Key Laboratory of
Wireless Communications, University of Electronic Science
and Technology of China, Chengdu 611731, China, and also with the Kash
Institute of Electronics and Information Industry, Kash 844000, China (e-mail: jqwang@uestc.edu.cn).
}

}
%

\maketitle

\begin{abstract}
This paper investigates a six-dimensional movable antenna (6DMA)-aided cell-free multi-user multiple-input multiple-output (MIMO) communication system. In this system, each distributed access point (AP) can flexibly adjust its array orientation and antenna positions to adapt to spatial channel variations and enhance communication performance. However, frequent antenna movements and centralized beamforming based on global instantaneous channel state information (CSI) sharing among APs entail extremely high signal processing delay and system overhead, which is difficult to be practically implemented in high-mobility scenarios with short channel coherence time. To address these practical implementation challenges and improve scalability, a two-timescale decentralized optimization framework is proposed in this paper to jointly design the beamformer, antenna positions, and array orientations. In the short timescale, each AP updates its receive beamformer based on local instantaneous CSI and global statistical CSI. In the long timescale, the central processing unit optimizes the antenna positions and array orientations at all APs based on global statistical CSI to maximize the ergodic sum rate of all users. The resulting optimization problem is non-convex and involves highly coupled variables, thus posing significant challenges for obtaining efficient solutions. To address this problem, a constrained stochastic successive convex approximation algorithm is developed. Numerical results demonstrate that the proposed 6DMA-aided cell-free system with decentralized beamforming significantly outperforms other antenna movement schemes with less flexibility and even achieves a performance comparable to that of the centralized beamforming benchmark.
\end{abstract}
\begin{IEEEkeywords}
Six-dimensional movable antenna (6DMA), cell-free MIMO, decentralized optimization, statistical CSI.
\end{IEEEkeywords}

\section{Introduction}
With the rapid evolution of wireless communications, the explosive growth of user and device connectivity has created an urgent demand for additional spectral resources. High-frequency bands such as millimeter-wave and terahertz have been regarded as key enablers for next-generation mobile communication networks. To compensate for severe path loss and improve signal coverage, it is necessary to employ large-scale antenna arrays and deploy \acp{ap} more densely. However, in the traditional cell-centric networks, the resulting high signal dimensionality introduced by large-scale arrays at the \ac{cpu} leads to substantial hardware cost and signal processing complexity for centralized computing. Meanwhile, densely deployed \acp{ap} in conventional cellular networks cause frequent handovers and strong inter-\ac{ap} interference, thereby degrading the performance of cell-edge users.

To address these challenges, cell-free massive \ac{mimo} has been proposed, in which a large number of distributed \acp{ap} collaboratively serve users without cell boundaries under the coordination of a \ac{cpu}. To further reduce the processing burden at the \ac{cpu}, cell-free systems can leverage \ac{lpu} of each \ac{ap} to locally process its received signals. Specifically, several studies have explored low-complexity decentralized beamforming schemes under different levels of \ac{csi} exchange \cite{emilbook,LorenzoTeamMMSE,LorenzoLoS,LorenzoTwoTimescale,qjshicell-free,zyhongtmmse}. In \cite{emilbook}, \ac{lmmse}-based beamforming methods were proposed in scenarios with no/statistical/full \ac{csi} sharing. In \cite{LorenzoTeamMMSE}, an optimal beamformer design rule that minimizes \ac{mse} under arbitrary \ac{csi}-sharing conditions was further developed. Building upon these \ac{lmmse}/\ac{mse}-based design strategies, the authors in \cite{LorenzoLoS,LorenzoTwoTimescale,qjshicell-free,zyhongtmmse,yczhangdistributedBF1,yczhangdistributedBF2} investigated decentralized architectures under specific scenarios, such as strong \ac{los} channels, satellite communications, uplink power minimization, \ac{csi} sharing among \acp{ap} based on a ring topology, and different levels of \ac{csi} sharing under channel uncertainty. However, in cell-free systems, the limited number of antennas at each AP makes it difficult to effectively mitigate inter-user interference. Moreover, most existing studies rely on \acp{fpa}, which cannot fully exploit the spatial \ac{dof} in wireless systems.

To address the limitations of \acp{fpa}, \ac{ma}, also known as \ac{fas}, has been proposed to improve wireless communication performance by flexibly adjusting antenna positions based on either instantaneous \ac{csi} or statistical \ac{csi} \cite{MA_tutorial,kkwfluid,lpzhuMA,wymaMAMIMO}. In \cite{lpzhuMA}, the authors first introduced the field-response-based channel model to analytically evaluate the performance gain of \ac{ma}-aided single-input single-output (SISO) systems. Building on this model, subsequent works \cite{wymaMAMIMO,lpzhumultiuser,lpzhuBeamforming,lpzhusatellite,lpzhuUAV,yczhangMA,xzyMUMIMO,xzyISAC} applied the MA technique to various scenarios under instantaneous CSI, where real-time antenna movement was leveraged to characterize the performance limits of various systems, including MIMO, multi-user, satellite, \ac{uav}, integrated sensing and communication (ISAC), and hybrid beamforming-aided communications systems. To alleviate the latency and power consumption induced by real-time antenna movement based on instantaneous CSI, discrete antenna movement strategy \cite{yfwu}, antenna movement trajectory optimization \cite{qllimindelay}, and the upper bound on energy efficiency \cite{jzdingdelay} were investigated to minimize movement delay or improve system energy efficiency. Furthermore, to avoid frequent antenna repositioning in high-mobility environments with relatively short channel coherence time, the authors in \cite{ypwuStatisCSI,yqye} optimized antenna positions based on statistical CSI to maximize MIMO capacity. In addition, two-timescale designs for beamforming and antenna position optimization were proposed in \cite{hgjietwotime,gystatistical} to improve the ergodic sum rate under both conventional Rician fading and statistical field-response-based channel models. Overall, these studies demonstrated that \ac{ma}-aided communication systems can achieve significant performance improvements under both instantaneous CSI and statistical CSI owing to their flexible spatial \ac{dof}.

More recently, \ac{6dma} has been proposed to further improve communication performance by jointly adjusting antenna positions and orientations, thereby fully exploiting spatial \ac{dof} for channel reconfiguration and interference suppression \cite{xdshao6dma,xdshaojsac,yczhang6dma,tsrenuav,hzwangpassive,xxiongestimation,xdshaonear}. In \cite{xdshao6dma}, the authors initially investigated \ac{6dma}-aided multi-user communications and optimized the positions and orientations of \ac{6dma} sub-arrays under \ac{los} channels according to the user distribution. Numerical results demonstrated that \ac{6dma} can provide substantial performance gains compared with the \ac{fpa} baseline. Subsequently, \cite{xdshaojsac,tsrenuav,yczhang6dma,hzwangpassive,xxiongestimation,xdshaonear} extended the \ac{6dma} to various scenarios, including discrete antenna position/orientation optimization, hybrid beamforming, \ac{uav} communications, intelligent reflecting surfaces (IRS)-aided passive beamforming, and near-field communications. Moreover, the authors in \cite{MA_tutorial,6dmaTutorial,xdshaotutorial,xzyChannelEstimation} introduced specific implementation architectures and practical considerations for \ac{6dma}, such as channel estimation accuracy, \ac{csi} acquisition timeliness, and hardware constraints. It is worth noting that most existing studies on \ac{6dma}-aided wireless networks rely on statistical \ac{csi} or user distribution to enable long-timescale antenna movement, aiming to achieve low-latency and cost-effective communication performance.

In particular, to enhance the communication performance in large-scale \ac{mimo} networks, prior works \cite{xdshaocellfree,xdshaocellfreeDMMSE,xypicellfree} have integrated \ac{6dma} into cell-free systems. The study in \cite{xypicellfree} adopted centralized beamforming and optimized antenna positions and orientations based on instantaneous CSI to support massive access and interference mitigation, whereas \cite{xdshaocellfree,xdshaocellfreeDMMSE} employed decentralized beamforming to reduce signal processing complexity. However, \cite{xdshaocellfree} considered an \ac{lmmse}-based beamformer design with no/full \ac{csi} sharing among \acp{ap}, which represent two extreme cases and do not capture the potential gains under practical \ac{csi} sharing conditions. In addition, only \ac{los} channels were considered for antenna movement optimization, which may result in significant performance degradation under practical multi-path fading channels that vary too fast for the antennas to adapt to due to limited antenna moving speeds. In contrast, the authors in \cite{xdshaocellfreeDMMSE} improved the decentralized beamforming design by introducing statistical \ac{csi} sharing among \acp{ap} but relying on real-time array reconfiguration, making it difficult to cope with the latency induced by frequent array rotations. Moreover, this work focused solely on array rotation, which restricts the available design \ac{dof} for reducing inter-user channel correlation, particularly in dense-user scenarios. Such limitation motivates a new design framework proposed in this paper for \ac{6dma}-aided cell-free systems.

In this work, we study a \ac{6dma}-aided uplink cell-free \ac{mumimo} communication system, where antennas at each \ac{ap} can move independently on an array panel, while the array panel can also rotate together with its antennas as a single unit. The main contributions are as follows.
\begin{itemize}
    \item To balance the antenna movement overhead and \ac{mimo} network performance, we develop a two-timescale decentralized optimization framework that maximizes the ergodic sum rate of all users. In the short timescale, each \ac{ap} designs its \ac{lmmse} receive beamformer using its local instantaneous \ac{csi} together with the statistical \ac{csi} shared among the \acp{ap}. In the long timescale, the \ac{cpu} optimizes antenna positions and array orientations based on global statistical \ac{csi}.
    \item To efficiently solve the non-convex long timescale antenna position and array orientation optimization problem, we reformulate a relaxed problem that treats the long-timescale parameters as independent optimization variables and propose a \ac{cssca} algorithm to solve the coupled variables in parallel.
    \item Simulation results show the performance gain of \ac{6dma} under both concentrated and uniformly distributed user scenarios. It is demonstrated that the proposed \ac{6dma}-aided cell-free system outperforms the conventional \ac{fpa} and other antenna movement schemes with less flexibility under different user distribution scenarios. Moreover, under specific conditions such as weaker inter-user channel correlations as well as sufficiently large movable regions or rotatable ranges, the proposed scheme can even surpass the performance of the centralized \ac{mmse} benchmark with \acp{fpa} that assumes full \ac{csi} sharing among all \acp{ap}. Moreover, the results indicate that the performance improvement becomes more pronounced under sparse user distributions, owing to the reduced channel correlation among users.

\end{itemize}

 The rest of this paper is organized as follows. Section II describes the system model and formulates the problem. Section III derives the optimal receive beamformer design rule over short timescale and then to solve the formulated long-timescale problem by the \ac{cssca} algorithm. Section IV presents numerical results and pertinent discussions. Finally, Section V concludes the paper. 

\emph{Notations:} Vectors and matrices are denoted in boldface with lower case and upper case, respectively. We use $\left(\cdot\right)^{\rm{T}}$, $\left(\cdot\right)^{*}$, and $\left(\cdot\right)^{\rm{H}}$ to denote the transpose, conjugate, and conjugate transpose, respectively. The notations $\mathbb{C}^{M\times N}$ and $\mathbb{R}^{M\times N}$ denote the set of $M\times N$ dimensional complex and real matrices, respectively. The notations $\mathbb{E}$ and $\mathbb{V}$ are the mathematical expectation and variance operators, respectively. We use ${\bf{I}}_N$ to denote the $N$-dimensional identity matrix, ${\bf{e}}_n\in\mathbb{R}^{N\times1}$ to denote the $N$-dimensional column vector with the $n$-th element being $1$ and all others being $0$, and ${\bf{A}}={\rm{blkdiag}}[{\bf{A}}_1,{\bf{A}}_2,\cdots,{\bf{A}}_N]$ to denote a block diagonal matrix. The notation $[f(n)]_{1\leq n\leq N}$ denotes a vector with the $n$-th element being $f(n)$, where $f(n)$ is a function w.r.t. $n$, $||{\bf{a}}||_2$ is the $2$-norm of vector ${\bf{a}}$. We use $\rm{arccos}(\cdot)$ to denote the arccosine function, $|b|$ and $\angle(b)$ are amplitude and phase of the complex number $b$, and ${\rm{Re}}\{b\}$ denotes the real part of $b$.

\section{System Model}
This section first presents the system model of the \ac{6dma}-aided cell-free network. A field-response-based statistical channel model is then adopted to characterize the phase and amplitude variations of quasi-static propagation paths at \acp{ap}, which are induced by antenna movement and array rotation \cite{lpzhuMA,gystatistical,xdshao6dma,xdshaojsac,xdshaocellfree,ypwuStatisCSI,SayeedFadChannel,PC_Heath}. Finally, the optimization problem is formulated to maximize the ergodic sum rate of all users, subject to practical antenna movement constraints.
  
\subsection{6DMA-aided Cell-free Architecture}
\begin{figure}
    \centering
    \includegraphics[width=0.8\linewidth]{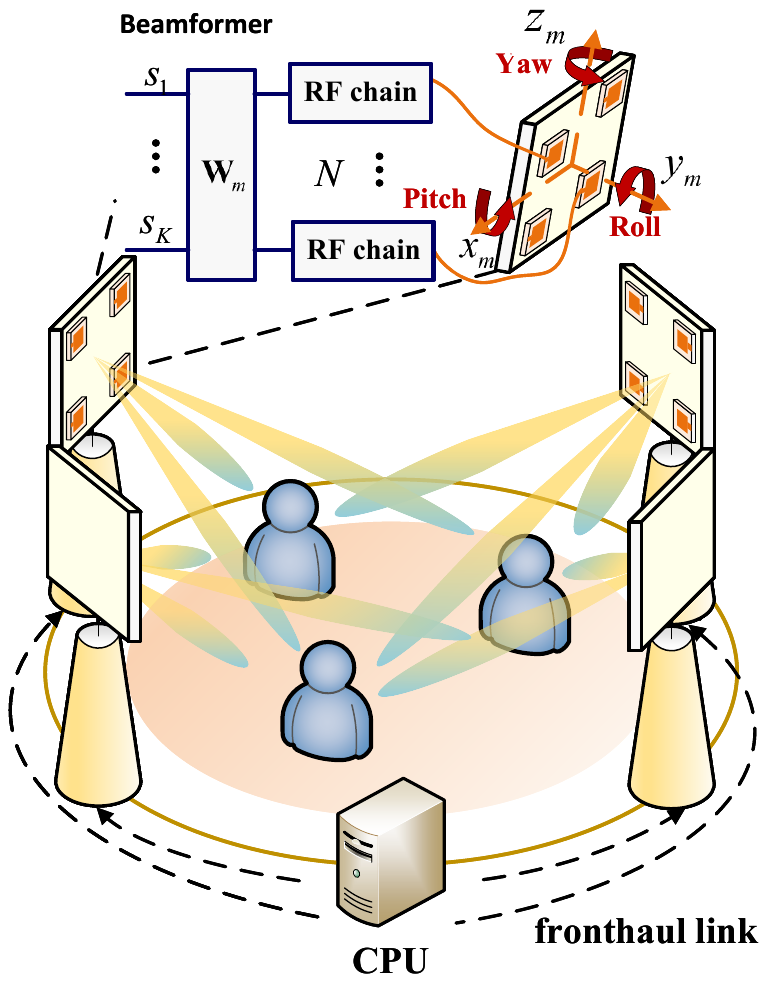}
    \caption{Illustration of the 6DMA-aided cell-free MU-MIMO system.}
    \label{system model}
\end{figure}

As illustrated in Fig.~\ref{system model}, we consider a \ac{6dma}-aided cell-free MU-MIMO uplink system, where $M$ multi-antenna \acp{ap}, each equipped with $N$ antennas, serve $K$ single-antenna users. A decentralized architecture is considered, where each \ac{ap} independently processes its received signals with an \ac{lpu} while exchanging signaling and payload data with the \ac{cpu} via a dedicated fronthaul link. The array of each \ac{ap} is capable of rotating around its geometric center. In addition, each antenna can move independently on the array plane driven by step motors. Due to practical implementation requirements, antenna movements and array rotations are restricted to a predefined range, and the inter-antenna spacing is no smaller than half a wavelength during antenna movement to avoid heavy mutual coupling.

 For the array of the $m$-th \ac{ap}, the red dotted line in Fig.~\ref{system model} denotes its \ac{lcs} ${x}_m$-${y}_m$-${z}_m$. The origin of this \ac{lcs} is fixed at the geometric center of the supporting plane, and its axes are aligned with the orientation of the array. The array orientation of the $m$-th \ac{ap} is denoted by ${{\bf{r}}_m=[\alpha_m,\beta_m,\gamma_m]^{\rm{T}}\in\mathcal{R}}$, where $\mathcal{R}$ denotes the feasible rotation region. Specifically, $\alpha_m$, $\beta_m$, and $\gamma_m$ represent the pitch, roll, and yaw angles of the $m$-th \ac{ap}’s rotation around the ${x}$-axis, ${y}$-axis, and ${z}$-axis with respect to (w.r.t.) its initial orientation, respectively. The positions of all antennas within the $m$-th \ac{ap}'s \ac{lcs} are collected as ${\bf{t}}_m=[{\bf{t}}_{m,1}^{\rm{T}},{\bf{t}}_{m,2}^{\rm{T}},\cdots,{\bf{t}}_{m,N}^{\rm{T}}]^{\rm{T}}\in\mathbb{R}^{3N\times 1}$, where ${\bf{t}}_{m,n}\in\mathcal{T}_{m,n}$ denotes the position of the $n$-th antenna at the $m$-th \ac{ap}, with $\mathcal{T}_{m,n}$ denoting the movable region of the $n$-th antenna.

\subsection{Statistical Channel Model}
Since the predefined movable regions of antennas at all \acp{ap} are much smaller than the propagation distance, the far-field assumption is adopted \cite{lpzhuMA,wymaMAMIMO,lpzhumultiuser,lpzhuBeamforming,lpzhusatellite,lpzhuUAV,yczhang6dma,xdshao6dma,xdshaojsac}. Antenna movement and array rotation change the three-dimensional (3D) antenna positions, boresight of antenna radiation patterns, and angles of arrival (AoAs) of the received signals. Under the multi-path propagation environment, the azimuth and elevation angles of the $l$-th received path w.r.t. the \ac{lcs} of the $m$-th \ac{ap}, before and after array rotation, are denoted by $(\phi_{l,m},\theta_{l,m})$ and $(\tilde{\phi}_{l,m},\tilde{\theta}_{l,m})$\footnote{For brevity, the subscripts associated with user indexes are omitted here, while the detailed channel model will be presented later.}, respectively.

\textit{1) Array Response Vectors:} We proceed to express the effective transformation from $(\phi,\theta)$ to $(\tilde{\phi},\tilde{\theta})$. Letting ${\bm{\Omega}}_{l,m}\triangleq[\phi_{l,m},\theta_{l,m}]$ with $\phi_{l,m}\in[0,2\pi)$ and $\theta_{l,m}\in[0,\pi]$, the normalized wave vector pointing to the wave direction $(\phi_{l,m},\theta_{l,m})$ is given by ${\bm{\rho}_{l,m}}\triangleq{\bm{\rho}({\bm{\Omega}}_{l,m})}=[\sin\theta_{l,m}\cos\phi_{l,m},\sin\theta_{l,m}\sin\phi_{l,m},\cos\theta_{l,m}]^{\rm{T}}$. We define ${\bf{R}}_z$, ${\bf{R}}_y$, and ${\bf{R}}_x$ as the \ac{lcs} transformation induced only by rotating w.r.t. $z$-axis, $y$-axis, and $x$-axis, respectively. The rotation matrix ${\bf{R}}$, which describes the array orientation transformation of any \ac{ap}, is expressed as
\vspace{-0.1cm}
\begin{equation}\label{rotation matrix}
\begin{aligned}
    &{\bf{R}}\left({\bf{r}}\right) = {\bf{R}}_z\left(\gamma\right){\bf{R}}_y\left(\beta\right){\bf{R}}_x\left(\alpha\right)\\
    &= \begin{bmatrix}
        c_{\gamma} & -s_{\gamma} & 0\\
        s_{\gamma} & c_{\gamma} & 0\\
        0 & 0 & 1
    \end{bmatrix}
    \begin{bmatrix}
        c_{\beta} & 0 & s_{\beta}\\
        0 & 1 & 0\\
        -s_{\beta} & 0 & c_{\beta} 
    \end{bmatrix}
    \begin{bmatrix}
        1 & 0 & 0\\
        0 & c_{\alpha} & -s_{\alpha}\\
        0 & s_{\alpha} & c_{\alpha}
    \end{bmatrix}\\
    &= \begin{bmatrix}
        c_{\beta}c_{\gamma} & s_{\alpha}s_{\beta}c_{\gamma}-c_{\alpha}s_{\gamma} & c_{\alpha}s_{\beta}c_{\gamma}+s_{\alpha}s_{\gamma} \\
        c_{\beta}s_{\gamma} & s_{\alpha}s_{\beta}s_{\gamma}+c_{\alpha}c_{\gamma} & c_{\alpha}s_{\beta}s_{\gamma}-s_{\alpha}c_{\gamma} \\
         -s_{\beta} & s_{\alpha}c_{\beta} & c_{\alpha}c_{\beta}
    \end{bmatrix},
\end{aligned}
\end{equation}
where $c_{\alpha}=\cos \alpha$ and $s_\alpha=\sin \alpha$ for concise. Letting ${\bf{R}}_m\triangleq {\bf{R}}({\bf{r}}_m)$, based on the geometric transformation, the direction vector of the $l$-th path w.r.t. the \ac{lcs} of the $m$-th \ac{ap} after array rotation is given by ${\tilde{\bm{\rho}}_{l,m}}\triangleq{\bm{\rho}(\tilde{\bm{\Omega}}_{l,m})}={\bf{R}}_m^{\rm{T}}\bm{\rho}_{l,m}$\cite{3gpp}, thus leading to
\begin{subequations}\label{coordinate trans}
    \begin{align}
        \tilde{\phi}_{l,m} &= \angle\left(
        [1,j,0]{\bf{R}}^{\rm{T}}_m{\bm{\rho}}_{l,m}\right)\label{phi_LCS},\\
        \tilde{\theta}_{l,m} &= \arccos\left(
        [0,0,1]{\bf{R}}^{\rm{T}}_m{\bm{\rho}}_{l,m}\right)\label{theta_LCS}.
    \end{align}
\end{subequations}
The array response vector of the $m$-th \ac{ap} for the $l$-th received path is then given by
\begin{equation}\label{steeringvector}
    {\bf{a}}\left({\bf{t}}_m,{\bf{r}}_m,{\bm{\Omega}}_{l,m}\right)=\left[e^{j\frac{2\pi}{\lambda}{\bm{\rho}}_{l,m}^{\rm{T}}{\bf{R}}_m{\bf{t}}_{m,n}}\right]_{1\leq n\leq N}^{\rm{T}},
\end{equation}
where $\lambda$ denotes the carrier wavelength. 

\begin{figure}
    \centering
    \includegraphics[width=0.8\linewidth]{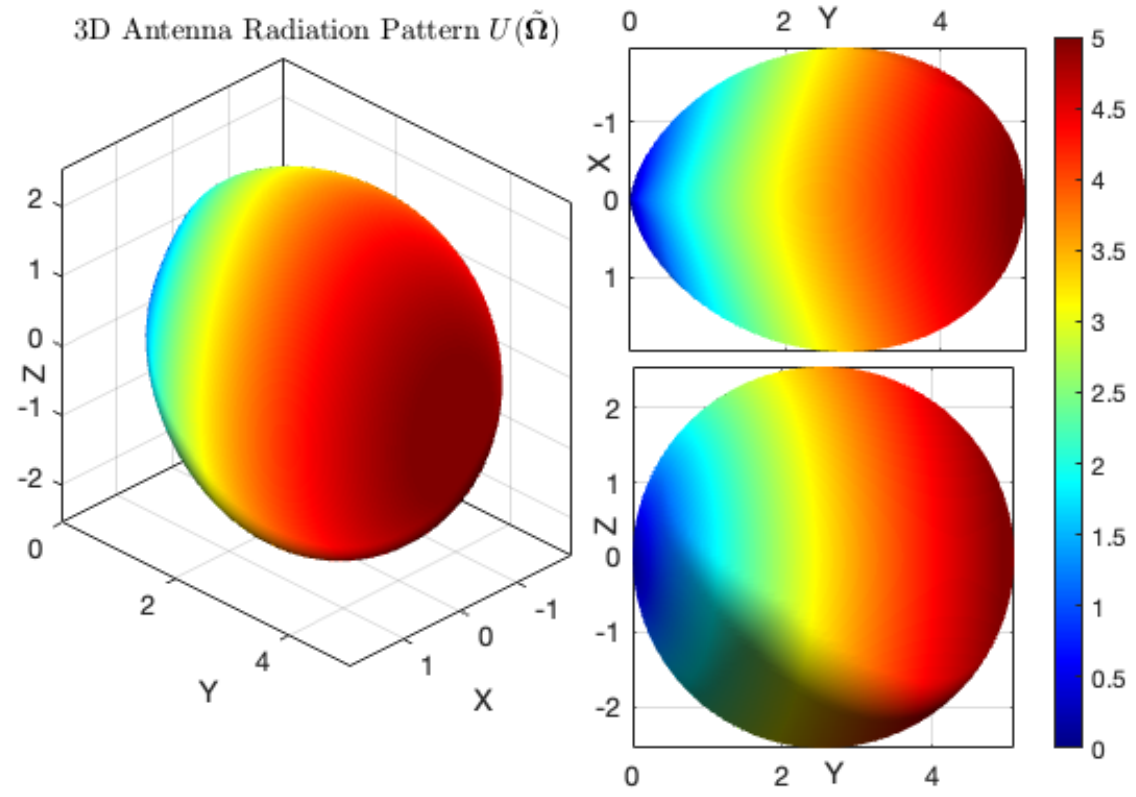}
    \caption{Illustration of the cosine radiation pattern \cite{antennapattern}.}
    \label{patternfig}
\end{figure}

\textit{2) Directional Radiation Pattern:} The array rotation also results in the variation of the boresight of each antenna's radiation pattern. As shown in Fig. \ref{patternfig}, for the commonly used cosine radiation pattern\cite{antennapattern}, the effective antenna gain for the $l$-th channel path at the $m$-th \ac{ap} is expressed as
\begin{equation}\label{radi pattern}
    U\left(\tilde{\bm{\Omega}}_{l,m}\right)=\left\{
    \begin{aligned}
        &{\frac{16}{\pi}\sin^2\tilde{\phi}_{l,m}\sin\tilde{\theta}_{l,m}},&&\tilde\phi_{l,m}\in\left[0,\pi\right),\\
        &0,&&{\rm{otherwise.}}
    \end{aligned}
    \right.
\end{equation}
Here, the radiated power is normalized by $\int_{\Omega}U(\phi,\theta){\rm{d}}\Omega=4\pi$ with $\Omega$ representing the unit sphere. We define $A_{\rm{E}}({\bf{r}}_{m},{\bm{\Omega}}_{l,m})\triangleq\sqrt{U(\tilde{\bm{\Omega}}_{l,m})}$ as the pattern coefficient along the propagation direction $(\phi_{l,m},\theta_{l,m})$.

To accurately evaluate the performance of \ac{6dma} in cell-free \ac{mimo} systems, it is essential to employ realistic channel models that capture key propagation characteristics. Most existing studies on \ac{6dma} or cell-free systems rely on deterministic channel models to assess performance limits~\cite{emilbook,xypicellfree,wymaMAMIMO,lpzhumultiuser,lpzhuBeamforming,yczhangMA,yczhang6dma}. {However, deterministic modeling, which assumes that the \ac{cpu} has complete knowledge of the \ac{csi} obtained through ray-tracing or other channel estimation approaches}, is challenging in practical deployments, as it entails substantial overhead for precise channel estimation and frequent antenna movement based on perfect \ac{csi}. Moreover, centralized processing, {in which all \acp{ap} collaboratively decode received signals through full \ac{csi} sharing}, requires extensive data and signal exchange between the \acp{ap} and the \ac{cpu}, further increasing the system burden. In high-mobility environments, the short channel coherence time is insufficient to support such intensive channel estimation, antenna movement, and \ac{ap} coordination. Motivated by these practical constraints, and to improve the applicability of \ac{6dma} in realistic communication scenarios, we adopt a statistical channel model following prior works~\cite{lpzhuMA,gystatistical,ypwuStatisCSI,SayeedFadChannel,PC_Heath,LorenzoTeamMMSE,LorenzoLoS,LorenzoTwoTimescale}.

Specifically, the field-response based channel model is considered. {In this model, it is assumed that the \ac{cpu} acquires the quasi-static AoAs for each \ac{ap}, while the small-scale fading coefficients vary rapidly due to environmental dynamics.} The channel response for narrow-band system between the $m$-th \ac{ap} and the $k$-th user is expressed as the sum of the responses of multiple channel paths:
\begin{equation}\label{channel_model}
\begin{aligned}
    {\bf{h}}_{k,m} &= \sum_{l=1}^{L_{k,m}}\psi_{k,l,m}A_{\rm{E}}\left({\bf{r}}_m,{\bm{\Omega}}_{k,l,m}\right){\bf{a}}\left({\bf{t}}_m,{\bf{r}}_m,{\bm{\Omega}}_{k,l,m}\right),
\end{aligned}
\end{equation}
where $L_{k,m}$ is the total number of received paths from the $k$-th user. $\psi_{k,l,m}\sim\mathcal{CN}(0,b_{k,l,m})$ denotes the fading coefficient of the $l$-th received paths between the $k$-th user and the $m$-th \ac{ap}, representing the small-scale fading effects associated with local scattering at the user side\cite{gystatistical,SayeedFadChannel,PC_Heath}, with $\{b_{k,l,m}\}$ being the average power. {In this decentralized architecture, each AP needs to access to both the local instantaneous CSI and the global statistical CSI. The local instantaneous CSI at the $m$-th AP is denoted as $\hat{\bf H}_{m} = [\hat{\bf h}_{1,m}, \hat{\bf h}_{2,m}, \dots, \hat{\bf h}_{K,m}]$, corresponding to specific estimated values of the fading coefficients $\{\{\psi_{k,l,m}\}_{l=1}^{L_{k,m}}\}_{k=1}^K$. In contrast, the global statistical CSI corresponds to the distribution of the random fading channel coefficients.}

\subsection{Signal Model}\label{ProbelmFormulate}
By collecting the channel vectors of all \acp{ap} in ${\bf{h}}_k=[{\bf{h}}_{k,1}^{\rm{T}},{\bf{h}}_{k,2}^{\rm{T}},\cdots,{\bf{h}}_{k,M}^{\rm{T}}]^{\rm{T}}$, the decoded signal of the $k$-th user at the \ac{cpu} is given by
\begin{equation}\label{received_signal}
\begin{aligned}
{{y}}_k=\sum_{m=1}^My_{k,m}={\bf{w}}^{\rm{H}}_k\left(\sum_{j=1}^K{\bf{h}}_j{{s}}_j+{\bf{n}}\right),
\end{aligned}
\end{equation}
where ${{s}}_k\sim\mathcal{CN}({{0}},1)$ is the transmit signal from the $k$-th user, $y_{k,m}$ denotes the signal decoded by the $m$-th \ac{ap} from the $k$-th user, ${\bf{w}}_k=[{\bf{w}}_{k,1}^{\rm{H}},{\bf{w}}_{k,2}^{\rm{H}},\cdots,{\bf{w}}_{k,M}^{\rm{H}}]^{\rm{H}}\in\mathbb{C}^{MN\times 1}$ and ${\bf{w}}_{k,m}$ are the receive beamformers for user $k$ at the \ac{cpu} and the $m$-th \ac{ap}, respectively. ${\bf{n}}\sim\mathcal{CN}({\bf{0}},\sigma^2{\bf{I}}_{MN})$ is the received Gaussian noise.

Based on the derived channel model in \eqref{channel_model} and signal model in \eqref{received_signal} , the achievable rate for the $k$-th user is given by
\begin{equation}
    R_k=\log_2\left(1+\frac{\left|{\bf{w}}_k^{\rm{H}}{\bf{h}}_k\right|^2}{\sum_{j\neq k}\left|{{\bf{w}}_k^{\rm{H}}{\bf{h}}_j}\right|^2+\sigma^2\left|\left|{\bf{w}}_k\right|\right|_2^2}\right).
\end{equation}

Several practical challenges arise when applying \ac{6dma} to cell-free networks. First, frequent \ac{csi} exchange among distributed \acp{ap} introduces considerable communication overhead, making centralized receive beamformer design based on full \ac{csi} difficult and thereby motivating decentralized optimization. Second, the relatively short channel coherence time makes it difficult to perform real-time antenna movement and forward the updates to all \acp{ap}. Frequent antenna adjustments may also lead to excessive processing latency and power consumption. Third, due to channel variations induced by antenna movement, the receive beamformer is coupled with antenna positions and array orientations, making joint optimization of the beamformer, antenna position, and array orientation highly challenging. To address these challenges, we propose a two-timescale design framework. In the short timescale, the receive beamformers at the $m$-th \ac{ap} are determined at its \ac{lpu} in closed form based on local instantaneous \ac{csi}. In the long timescale, antenna movement at each \ac{ap} is performed using slow-varying statistical \ac{csi}, and the optimized antenna positions and array orientations are forwarded to all \acp{ap}.

Denote ${\bf{t}}=[{\bf{t}}_1^{\rm{T}},{\bf{t}}_2^{\rm{T}},\cdots,{\bf{t}}_M^{\rm{T}}]^{\rm{T}}$, ${\bf{r}}=[{\bf{r}}_1^{\rm{T}},{\bf{r}}_2^{\rm{T}},\cdots,{\bf{r}}_M^{\rm{T}}]^{\rm{T}}$, and ${\bf{W}}=[{\bf{w}}_1,{\bf{w}}_2,\cdots,{\bf{w}}_K]$ for notation simplicity. To maximize the ergodic sum rate of all users, the two-timescale optimization problem that jointly designs the receive beamformer ${\bf{W}}$, the array orientations ${\bf{r}}$, and the antenna positions ${\bf{t}}$ of all \acp{ap} is given by
\begin{subequations}\label{optimization_problem}
\begin{align}
    \max_{{\bf{t}},{\bf{r}}}&~{\mathbb{E}}_{\bf H}\left[\max_{{\bf W}}\sum_{k=1}^KR_k\right]\label{obj}\\
    {\rm{s.t.}}&~{\bf{t}}_{m,n}\in\mathcal{T}_{m,n},~\forall m,n,\label{position_constraint}\\
    &~{\bf{r}}_{m}\in\mathcal{R},~\forall m,\label{rotation_constraint}
\end{align}
\end{subequations}
where $\mathcal{R}=\{[\alpha,\beta,\gamma]^{\rm{T}}|\alpha\in [\alpha^{\rm{L}},\alpha^{\rm{U}}],\beta\in [\beta^{\rm{L}},\beta^{\rm{U}}],\gamma\in [\gamma^{\rm{L}},\gamma^{\rm{U}}]\}$ and $\mathcal{T}_{m,n}=\{[x,y,z]^{\rm{T}}|x\in [x_{m,n}^{\rm{L}},x_{m,n}^{\rm{U}}],y\in [y_{m,n}^{\rm{L}},y_{m,n}^{\rm{U}}],z\in [z_{m,n}^{\rm{L}},z_{m,n}^{\rm{U}}]\}$ denote the predefined rotatable and movable ranges, respectively. The variables with superscripts ${\rm{L}}$ and ${\rm{U}}$ denote the lower and upper bounds on the corresponding variables, respectively. {In the rest of the paper, the expectation operator ${\mathbb E}[\cdot]$ is applied to the random channel matrix ${\bf H}$ unless stated otherwise. It is important to note that optimizing the beamformer ${\bf W}$ is a distributed process, where each AP optimizes its local receive beamformer based on the available CSI. This necessitates an update rule for the beamformer at each AP to enhance the overall sum rate. Furthermore, due to the expectation over ${\bf H}$, determining the optimal antenna positions and array orientations is challenging.}

\begin{remark}\label{remark1}
Due to the lack of instantaneous \ac{csi} from other \acp{ap}, the receive beamformer at each \ac{ap} should be updated in the short timescale by jointly exploiting the global statistical CSI and the local instantaneous CSI, as discussed in Section~\ref{ProbelmFormulate}. Consequently, in the formulation of the ergodic sum rate optimization problem \eqref{optimization_problem}, the receive beamformer, ${\bf W}$, should be regarded as a mapping from the available \ac{csi} to the receive beamforming matrix rather than as a fixed solution.
\end{remark}

\begin{figure}
    \centering
    \includegraphics[width=1.0\linewidth]{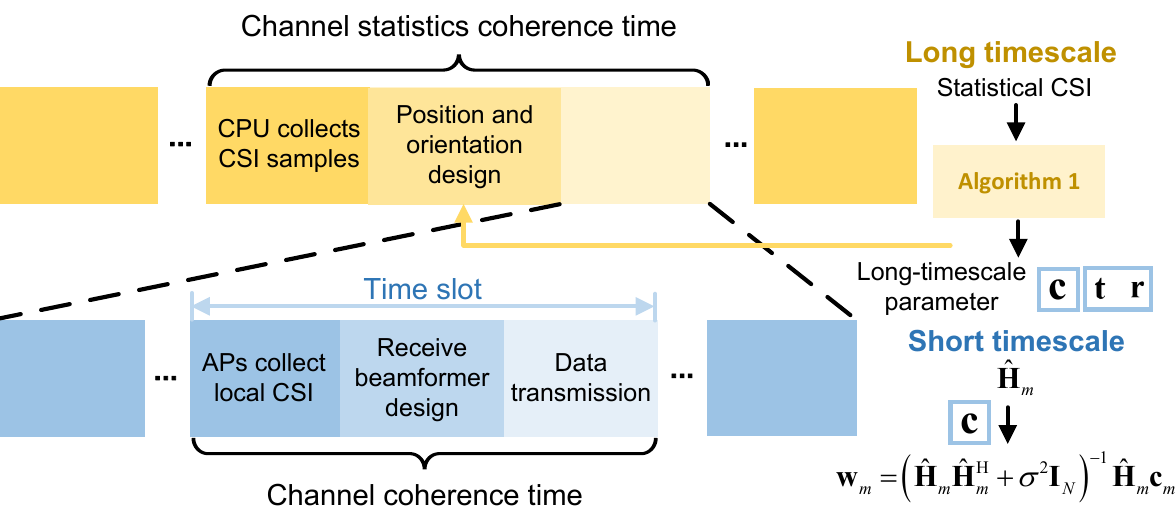}
    \caption{Illustration of the two-timescale decentralized design framework, including the short-timescale workflow at each \ac{ap} and the long-timescale workflow at the \ac{cpu}.}
    \label{timescale}
\end{figure}

\section{Two-timescale Decentralized Design}

In this section, we aim to solve problem \eqref{optimization_problem}. Specifically, a two-timescale design is proposed to balance practical feasibility and ergodic sum-rate performance. As illustrated in Fig. \ref{timescale}, the long-timescale parameter is treated as a constant within the short-timescale optimization, which are computed at the \ac{cpu} and then distributed to all \acp{ap} for local processing. To enable each \ac{ap} to design its short-timescale receive beamformer in closed form based on local instantaneous \ac{csi} and global statistical \ac{csi}, the objective function is first approximated using the well-known \ac{uatf} lower bound \cite{uatfbound} for improved tractability. Subsequently, inspired by the \ac{FP} framework \cite{FP}, the optimal receive beamformer is derived w.r.t. both statistical and local instantaneous \ac{csi}. In the long timescale, the \ac{cssca} algorithm is employed to jointly optimize the antenna positions and array orientations in parallel to maximize the ergodic sum rate of all users.
\subsection{Short-timescale Receive Beamformer Design}
\subsubsection{FP-Based UatF Bound Optimization}
To efficiently derive a closed-form expression of the receive beamformer at each AP, we optimize the more tractable \ac{uatf} lower bound on the ergodic achievable rate for each AP, which is given by \cite{uatfbound}

\begin{subequations}
\begin{align}
&\sum_{k=1}^K{\mathbb{E}}\left[R_k\right]\geq \sum_{k=1}^KR_k^{\rm{UatF}}=\sum_{k=1}^K\log\left(1+{\rm{SINR}}_k^{\rm{UatF}}\right),\label{UatF_Bound}\\
&{\rm{SINR}}_k^{\rm{UatF}}=\frac{|\mathbb{E}[{\bf{w}}_k^{\rm{H}}{\bf{h}}_k]|^2}{\sum\limits_{j\neq k}\mathbb{E}[|{\bf{w}}_k^{\rm{H}}{\bf{h}}_j|^2]+\mathbb{V}[{\bf{w}}_k^{\rm{H}}{\bf{h}}_k]+\sigma^2\mathbb{E}[||{\bf{w}}_k||_2^2]}.
\end{align}
\end{subequations}

Then, during the short timescale, given antenna positions ${\bf t}$ and array orientations ${\bf r}$, the subproblem for optimizing receive beamformers can be written as
\begin{equation}\label{optimization_problem_UatF}
    \max_{{\bf W}}~\sum_{k=1}^KR_k^{\rm UatF}.
\end{equation}
According to the \ac{FP} framework \cite{FP}, the problem in \eqref{optimization_problem_UatF} can be equivalently reformulated as

\begin{equation}\label{FPformulate}
        \min_{\substack{
            {\bf{W}},{\bm{\vartheta}},\,{\bm{\varpi}}
        }}
        \sum_{k=1}^K\left(1+\vartheta_k\right)\,{\mathbb{E}}\!\left[e_k\left(\varpi_k\right)\right]
        +\vartheta_k-\log\left(1+\vartheta_k\right),
\end{equation}
where ${\bm{\vartheta}}=\{\vartheta_k\}_{k=1}^{K}$ and ${\bm{\varpi}}=\{\varpi_k\}_{k=1}^{K}$ are both a set of the slack variables. $\mathbb{E}[e_k(\varpi_k)]$ is given by
\begin{equation}\label{MSE}
\begin{aligned}{\mathbb{E}}\left[e_k\left(\varpi_k\right)\right]&=\left|\varpi_k\right|^2\Bigg(\sum\limits_{j\neq k}\mathbb{E}\left[\left|{\bf{w}}_k^{\rm{H}}{\bf{h}}_j\right|^2\right]+\mathbb{V}\left[{\bf{w}}_k^{\rm{H}}{\bf{h}}_k\right]\\&\quad+\sigma^2\mathbb{E}\left[\left|\left|{\bf{w}}_k\right|\right|_2^2\right]\Bigg)-2{\rm{Re}}\left\{\varpi_k^*\mathbb{E}\left[{\bf{w}}_k^{\rm{H}}{\bf{h}}_k\right]\right\}\\
   &\stackrel{\left(a\right)}{=}\mathbb{E}\left[\left|\left|\varpi_k{\bf{H}}^{\rm{H}}{\bf{w}}_k-{\bf{e}}_k\right|\right|_2^2+\sigma^2\left|\left|\varpi_k{\bf{w}}_k\right|\right|_2^2\right],
\end{aligned}
\end{equation}
where ${\bf H}=[{\bf h}_1,{\bf h}_2,\cdots,{\bf h}_K]$. The equality $(a)$ is derived using the identity $\mathbb{V}[{\bf{w}}^{\rm H}{\bf{h}}] = \mathbb{E}[|{\bf{w}}^{\rm H}{\bf{h}}|^2] - |\mathbb{E}[{\bf{w}}^{\rm H}{\bf{h}}]|^2$.
\subsubsection{Receive Beamformer Design}
In the short timescale, the global statistical \ac{csi} is assumed to be shared among the \acp{ap}. With the antenna positions and array orientations optimized over the long timescale being fixed, the receive beamformer at the $m$-th AP is obtained by maximizing the ergodic sum rate in \eqref{optimization_problem_UatF} based on the available \ac{csi} $\hat{\bf{H}}_m$. From problem \eqref{FPformulate}, it can be observed that the receive beamformer for the $k$-th user’s signal is determined solely by the $k$-th \ac{mse} term ${\mathbb{E}}[e_k(\varpi_k)]$. By defining $\tilde{\bf{w}}_k\triangleq[\tilde{\bf{w}}_{k,1}^{\rm{H}},\tilde{\bf{w}}_{k,2}^{\rm{H}},\cdots,\tilde{\bf{w}}_{k,M}^{\rm{H}}]^{\rm{H}}=\varpi_k^{\rm opt}{\bf{w}}_k$ and $\tilde{\bf W}_m=[\tilde{\bf{w}}_{1,m},\tilde{\bf{w}}_{2,m},\cdots,\tilde{\bf{w}}_{K,m}]$, the short-timescale optimization problem for the $m$-th AP, based on available \ac{csi} $\hat{\bf H}_m$, can be equivalently simplified from problem \eqref{FPformulate} as

\begin{equation}\label{ShortTimescale}
\begin{aligned}
        \min_{{\tilde{\bf{W}}_{m}}}
        \sum_{k=1}^K\mathbb{E}&\left[\Bigg|\Bigg|\sum_{i\neq m}{\mathbf{H}}_i^{\rm{H}}\tilde{\mathbf{w}}_{k,i}+{\mathbf{H}}_m^{\rm{H}}\tilde{\bf{w}}_{k,m}-{\bf{e}}_k\Bigg|\Bigg|_2^2\right.\\
        &+\sigma^2\left.\left.\left(\sum_{i\neq m}\left|\left|\tilde{\mathbf{w}}_{k,i}\right|\right|_2^2+\left|\left|\tilde{\bf{w}}_{k,m}\right|\right|_2^2\right)\right|\hat{\bf{H}}_m\right].
\end{aligned}
\end{equation}
Notably, as stated in {\bf{Remark} \ref{remark1}}, the receive beamformers at the other \acp{ap} should be treated as functions of random fading channels. Since the objective function in \eqref{ShortTimescale} is convex and continuously differentiable w.r.t. $\tilde{\bf{w}}_{k,m}$, the optimal solution can be characterized by the first-order optimality condition, i.e.,
\begin{equation}\label{optimalitycondition}
    \hat{\bf{H}}_m\left(\sum_{i\neq m}\mathbb{E}\left[{\mathbf{H}}_i^{\rm{H}}\tilde{\mathbf{w}}_{k,i}\right]+\hat{\bf{H}}_m^{\rm{H}}\tilde{\bf{w}}_{k,m}-{\bf{e}}_k\right)+\sigma^2\tilde{\bf{w}}_{k,m}=0.
\end{equation}
Hence, letting $\hat{\bf{G}}_m=(\hat{\bf{H}}_m\hat{\bf{H}}_m^{\rm{H}}+\sigma^2{\bf{I}}_N)^{-1}\hat{\bf{H}}_m$, the optimal value of 
$\tilde{\bf{w}}_{k,m}$ in problem \eqref{ShortTimescale} is derived as
\begin{equation}\label{receive beamformeropt}
    \tilde{\bf{w}}_{k,m}^{\rm{opt}}\left(\hat{\bf{H}}_m\right)=\hat{\bf{G}}_m\left({\bf{e}}_k-\sum_{i\neq m}\mathbb{E}\left[{\mathbf{H}}_i^{\rm{H}}\tilde{\mathbf{w}}_{k,i}\right]\right).
\end{equation}
\begin{remark}
It should be emphasized that although \eqref{receive beamformeropt} yields the real-time per-AP optimal receive beamformer under available \ac{csi}, its computation depends on the statistical distributions of the channels and associated receive beamformers at the other \acp{ap}. Due to this interdependency, \eqref{receive beamformeropt} effectively introduces an iteratively update rule analogous to alternating optimization procedure, in which the $m$-th \ac{ap}'s receive beamformer is optimized while the statistical distributions of the receive beamformers at the remaining \acp{ap} are kept fixed. Therefore, it is useful to establish that this iterative procedure converges to a stationary point and to derive a closed-form expression for that stationary point. From the perspective of team theory, the authors in \cite{LorenzoTeamMMSE} rigorously proved that when all APs adopt \eqref{receive beamformeropt} to design their receive beamformers, the resulting solution is the unique global optimum of the original problem in \eqref{optimization_problem_UatF}. {Furthermore, the short-timescale optimization problem in \eqref{ShortTimescale} can be generalized to incorporate any level of available \ac{csi}. Specifically, when the available \ac{csi} corresponds to the global instantaneous \ac{csi} $\hat{\bf H}$, the problem reduces to a centralized beamformer design.}
\end{remark}

To further eliminate the interdependency among the receive beamformers of different \acp{ap} in \eqref{receive beamformeropt}, a closed-form expression for the \ac{lmmse} beamformer satisfying \eqref{receive beamformeropt} is given by 
\begin{equation}\label{closeformreceive beamformer}
    \tilde{\bf{w}}_{k,m}^{\rm{opt}}\left(\hat{\bf{H}}_m\right)=\hat{\bf{G}}_m{\bf{c}}_{k,m},~\forall m,
\end{equation}
where ${\bf{c}}_{k}=[{\bf{c}}_{k,1}^{\rm{T}},{\bf{c}}_{k,2}^{\rm{T}},\cdots,{\bf{c}}_{k,M}^{\rm{T}}]^{\rm{T}}\in\mathbb{C}^{KM\times 1}$ represents a set of long-timescale parameters that facilitate all \acp{ap} to locally process the received signals from the $k$-th user. These parameters are determined by the statistical \ac{csi} and calculated as \cite{LorenzoTeamMMSE}
\begin{equation}\label{longtimevariable}
    {\bf{c}}_k=\left({\rm{blkdiag}}\left({\bf{U}}-{\mathbf{V}}\right)+{\bf{U}}^{\rm{T}}{\bf{V}}\right)^{-1}{\bf{U}}^{\rm{T}}{\bf{e}}_k,
\end{equation}
where ${\bf{V}}={\mathbb{E}}\{[{\mathbf{H}}_1^{\rm{H}}{\mathbf{G}}_1,{\mathbf{H}}_2^{\rm{H}}{\mathbf{G}}_2,\cdots,{\mathbf{H}}_M^{\rm{H}}{\mathbf{G}}_M]\}\in{\mathbb{C}}^{K\times KM}$ and ${\bf{U}}=[{\bf{I}}_K,{\bf{I}}_K,\cdots,{\bf{I}}_K]\in{\mathbb{R}}^{K\times KM}$. The proof of the existence and uniqueness of ${\bf{c}}_{k}$ is provided in Appendix~\ref{appendixa}. By substituting \eqref{closeformreceive beamformer} and \eqref{longtimevariable} into \eqref{receive beamformeropt}, it follows directly that \eqref{closeformreceive beamformer} constitutes an optimal solution for \eqref{receive beamformeropt}. Finally, since problem \eqref{FPformulate} w.r.t. the receive beamformers is an unconstrained optimization problem and the fact of $R_k^{\rm{UatF}}(\tilde{\bf{w}}_k)=R_k^{\rm{UatF}}({\bf{w}}_k)$, both ${\bf w}_k$ and $\tilde{\bf w}_k$ are locally optimal beamformers, resulting in the same ergodic sum rate. Therefore, we can directly update the beamformer using \eqref{closeformreceive beamformer}.

\subsection{Long-timescale Antenna Reconfiguration}
In the long-timescale design, the receive beamformers based on the instantaneous CSI are regarded as functions of random \ac{csi}. By substituting the optimized receive beamformer in \eqref{closeformreceive beamformer} into \eqref{optimization_problem} and defining ${\bf G}={\rm blkdiag}[{\bf G}_1,{\bf G}_2,\cdots,{\bf G}_M]$, the corresponding long-timescale subproblem w.r.t antenna positions and array orientations is formulated as
\begin{subequations}\label{longtimepro}
\begin{align}
    \max_{{\bf{t}},{\bf{r}}}&~\mathbb{E}\scalebox{1.5}{$\Bigg[$}\underbrace{
\sum_{k=1}^K\log_2\left(1+\frac{\left|{\bf{c}}_{k}^{\rm{H}}{\mathbf{G}}^{\rm{H}}{\mathbf{h}}_{k}\right|^2}{\sum\limits_{j\neq k}\left|{\bf{c}}_{k}^{\rm{H}}{\mathbf{G}}^{\rm{H}}{\mathbf{h}}_{j}\right|^2+\sigma^2\left|\left|{\mathbf{G}}{\bf{c}}_k\right|\right|^2_2}\right)}_{g\left\{{\mathbf{H}}\left({\bf{t}},{\bf{r}}\right),{\bf{c}}\right\}}\scalebox{1.5}{$\Bigg]$}\label{longtimeobj}\\
    {\rm{s.t.}}&~\eqref{position_constraint},\eqref{rotation_constraint},
\end{align}
\end{subequations}
where ${\bf{c}}=[{\bf{c}}_1^{\rm{T}},{\bf{c}}_2^{\rm{T}},\cdots,{\bf{c}}_K^{\rm{T}}]^{\rm{T}}\in\mathbb{C}^{K^2M\times 1}$ denotes the long-timescale parameter, with ${\bf{c}}_k$ given by \eqref{longtimevariable}. It should be emphasized that the optimal long-timescale parameter ${\bf{c}}$ given in \eqref{longtimevariable} is determined by the statistical \ac{csi}, and thus implicitly depends on the antenna positions ${\bf t}$ and array orientations ${\bf r}$. This interdependency induces strong coupling among ${\bf t}$, ${\bf r}$, and ${\bf c}$ in the objective function 
\eqref{longtimeobj}, making the long-timescale optimization problem difficult to solve directly. 
To address this coupling, the \ac{saa} is commonly employed to approximate the long-timescale parameter and the ergodic sum rate of all users via averaging over multiple random channel realizations\cite{gystatistical,xdshao6dma,xypicellfree,xdshaocellfree}. However, achieving sufficiently accurate approximation requires extensive sampling, which substantially increases the computational complexity in large-scale cell-free \ac{mimo} systems.

\begin{figure*}[b]
    \normalsize
    \setcounter{equation}{23}
    \hrulefill
    \begin{equation}
        \label{partial_c}
        \nabla_{{\bf{c}}_k}g\left({\bf t},{\bf r},{\bf c},{\bm \Psi}\right)=\frac{1}{\ln 2}\left(\frac{{\mathbf{G}}^{\rm{H}}\sum\limits_{j=1}^K{\mathbf{h}}_{j}{\mathbf{h}}_{j}^{\rm H}{\mathbf{G}}{\bf{c}}_{k}+\sigma^2{\mathbf{G}}^{\rm H}{\mathbf{G}}{\bf{c}}_{k}}{\sum\limits_{j=1}^K\left|{\bf{c}}_{k}^{\rm{H}}{\mathbf{G}}^{\rm{H}}{\mathbf{h}}_{j}\right|^2+\sigma^2\left|\left|{\mathbf{G}}{\bf{c}}_k\right|\right|^2_2}-\frac{{\mathbf{G}}^{\rm{H}}\sum\limits_{j\neq k}{\mathbf{h}}_{j}{\mathbf{h}}_{j}^{\rm H}{\mathbf{G}}{\bf{c}}_{k}+\sigma^2{\mathbf{G}}^{\rm H}{\mathbf{G}}{\bf{c}}_{k}}{\sum\limits_{j\neq k}\left|{\bf{c}}_{k}^{\rm{H}}{\mathbf{G}}^{\rm{H}}{\mathbf{h}}_{j}\right|^2+\sigma^2\left|\left|{\mathbf{G}}{\bf{c}}_k\right|\right|^2_2}\right),~1\leq k\leq K.
    \end{equation}
\setcounter{equation}{18}
\end{figure*}

Given this challenge, it is necessary to develop an efficient optimization framework that can jointly address the interdependency among variables and reduce the computational burden. {Although the optimal long-timescale parameter ${\mathbf c}$ is formally characterized in \eqref{longtimevariable}, it involves an expectation operator embedded in a nonlinear function, which renders the problem analytically intractable. It is worth noting that the optimal ${\mathbf c}$ is independent of the instantaneous \ac{csi}. This observation naturally motivates a centralized computation of ${\mathbf c}$ at the \ac{cpu}, followed by distributing it to all \acp{ap}. At the \ac{ap} level, each \ac{ap} then determines its receive beamformer by combining the locally available instantaneous \ac{csi} with the provided long-term parameter ${\mathbf c}$. Therefore, to render the problem in \eqref{longtimepro} tractable, we treat ${\mathbf c}$ as an auxiliary variable at the \ac{cpu}. This relaxation leads to a computationally feasible yet suboptimal solution and converts the original problem in \eqref{longtimepro} into the following subproblem:}
\begin{equation}\label{longtermparaformulate}
\begin{aligned}
\max_{{\bf{t}},{\bf{r}},{\bf{c}}}&~\mathbb{E}\left[g\left\{{\mathbf{H}}\left({\bf{t}},{\bf{r}}\right),{\bf{c}}\right\}\right]\\
{\rm{s.t.}}&~\eqref{position_constraint},\eqref{rotation_constraint}.
\end{aligned}
\end{equation}
The coupled variables and expectation still lead to stochasticity and non-concavity of problem \eqref{longtermparaformulate}. As discussed above, to further decouple and then efficiently optimize these variables, the \ac{cssca} framework is employed. In this framework, the original problem is iteratively addressed through a series of concave surrogate functions. In particular, the surrogate-function structure enables natural decoupling among different optimization variables, allowing them to be separately updated within each iteration. This facilitates online optimization while maintaining fast convergence and high computational efficiency \cite{LAcssca}, which will be detailed later.

The \ac{cssca} procedure is described as follows. Recalling the channel model in \eqref{channel_model}, it is noteworthy that only ${\bf{\Psi}}\triangleq\{\{\{\psi_{k,l,m}\}_{m=1}^M\}_{l=1}^{L_k}\}_{k=1}^K$ are random variables. In the $s$-th iteration, a set of channel realization ${\bf{\Psi}}^s$ is realized to update the surrogate function $\bar{f}^{s}({\bf{t}}^{s},{\bf{r}}^{s},{\bm\Psi}^s)$, which can be regarded as a concave approximation of the objective function in \eqref{longtermparaformulate} to calculate the long-timescale parameter, antenna positions, and array orientations in the next iteration.

Specifically, we adopt the smooth surrogate function construction proposed in \cite{LAcssca}, which can be viewed as a convex approximation of $g\{{\bf H}({\bf t},{\bf r}),{\bf c}\}$ and is constructed as
\begin{equation}\label{surrogate}
\begin{aligned}
    \bar{f}^{s}\left({\bf{t}},{\bf{r}},{\bf{c}},{\bm{\Psi}}^s\right)&=f^{s}+{{f}}_{\bf{t}}^{s}\left({\bf{t}}\right)+{{f}}_{\bf{r}}^{s}\left({\bf{r}}\right)+{{f}}_{\bf{c}}^{s}\left({\bf{c}}\right),
\end{aligned}
\end{equation}
where $f^s$ is an iterative average measure of the objective function in \eqref{longtermparaformulate}, which is given by
\begin{equation}
    f^s=\frac{1}{s}\sum_{u=1}^sg\left\{{\bf{H}}\left({\bf{t}}^{s},{\bf{r}}^{s},{\bm{\Psi}}^u\right),{\bf{c}}^{s}\right\}.
\end{equation}
${{f}}_{\bf{t}}^{s}\left({\bf{t}}\right)$, ${{f}}_{\bf{r}}^{s}\left({\bf{r}}\right)$, and ${{f}}_{\bf{c}}^{s}\left({\bf{c}}\right)$ are given by
\begin{equation}\label{surrogate function linear}
\begin{aligned}
{f}^{s}_{\bf v}\left({\bf{v}}\right)={\rm Re}\left\{\left({\bf{f}}_{\bf{v}}^{s}\right)^{\rm{H}}\left({\bf{v}}-{\bf{v}}^s\right)+\tau_{\bf{v}}\left|\left|{\bf{v}}-{\bf{v}}^s\right|\right|_2^2\right\},\\{\bf v}\in\{{\bf t},{\bf r},{\bf c}\},
\end{aligned}
\end{equation}
where $\{\tau_{\bf{v}}\}_{{\bf v}\in\{{\bf t},{\bf r},{\bf c}\}}$ are negative constants. The first and second terms in \eqref{surrogate function linear} are derived from the linearization of the sampled objective function and the proximal regularization term, respectively. For notation simplicity, denote $g({\bf t},{\bf r},{\bf c},{\bm \Psi})\triangleq g\left\{{\bf{H}}\left({\bf{t}},{\bf{r}},{\bm{\Psi}}\right),{\bf{c}}\right\}$. The term ${\bf{f}}_{\bf{v}}^{s}$ approximates the first-order gradient of the objective function in \eqref{longtermparaformulate} w.r.t. the variable ${\bf v}$ and can be calculated recursively as
\begin{equation}
    {\bf{f}}_{\bf{v}}^{s}=\left(1-\rho^s\right){\bf{f}}_{\bf{v}}^{s-1}+\rho^s\nabla_{\bf{v}}g\left({\bf t}^s,{\bf r}^s,{\bf c}^s,{\bm \Psi}^s\right),{\bf v}\in\{{\bf t},{\bf r},{\bf c}\},\label{gradient}
\end{equation}
where $f^{-1}\triangleq0$ and $\{\rho^s\in(0,1]\}$ is a decreasing sequence such that $\lim\limits_{s\rightarrow\infty}\rho^s=0$, $\lim\limits_{s\rightarrow\infty}\sum_{s}\rho^s\rightarrow\infty$, and $\lim\limits_{s\rightarrow\infty}\sum_s(\rho^s)^2\leq\infty$. These conditions confirm the convergence of the proposed \ac{cssca} procedure. The gradient of $g({\bf t},{\bf r},{\bf c},{\bm \Psi})$ w.r.t. ${\bf c}$ can be derived in closed form, as shown in \eqref{partial_c} at the bottom of the previous page. Moreover, the gradients w.r.t. ${\bf t}$ and ${\bf r}$ are computed numerically. Specifically, the gradient w.r.t. ${\bf r}$ is given by
\begin{equation}
\setcounter{equation}{25}
\begin{aligned}
    &\left[\nabla_{\bf{r}}g\left({\bf t}^s,{\bf r}^s,{\bf c}^s,{\bm \Psi}^s\right)\right]_i=\\
    &\lim_{\varepsilon\rightarrow0}\frac{g\left({\bf t}^s,{\bf r}^s+\varepsilon{\bf e}_i,{\bf c}^s,{\bm \Psi}^s\right)-g\left({\bf t}^s,{\bf r}^s,{\bf c}^s,{\bm \Psi}^s\right)}{\varepsilon},1\leq i\leq 3M,
\end{aligned}
\end{equation}
and the gradient w.r.t. ${\bf t}$ can be obtained in a similar manner.
Then, according to the \ac{cssca} procedure \cite{LAcssca}, the optimization problem in the $s$-th iteration is recast into
\begin{subequations}\label{CSSCAoptimization}
\begin{align}
    \max_{{\bf{t}},{\bf{r}},{\bf c}}&~\bar{f}^{s}\left({\bf{t}},{\bf{r}},{\bf{c}},{\bm{\Psi}}^s\right)\\
    {\rm{s.t.}}&~\eqref{position_constraint},\eqref{rotation_constraint}.
\end{align}
\end{subequations}
Given that the decoupled construction of surrogate functions, problem \eqref{CSSCAoptimization} can be equivalently decoupled into three subproblems. {Since the surrogate objective functions in \eqref{surrogate function linear} are convex quadratic and each subproblem is either unconstrained or subject to linear constraints, the unique solution for each can be determined in closed-form. Specifically, the optimal solution $\{\bar{\bf{t}}^{s},\bar{\bf{r}}^{s},\bar{\bf{c}}^{s}\}$ is derived by projecting the unconstrained stationary point of the quadratic function onto its respective feasible region, i.e.,}
\begin{equation}\label{subposition}
    \bar{\bf{t}}^{s}=\underset{\left\{{\bf{t}}_{m,n}\in{\mathcal T}_{m,n}\right\}_{\forall m,n}}{\rm argmax}\bar{f}^{s}_{\bf t}\left({\bf{t}}\right)=\mathcal{B}_{\rm{t}}\left\{{\bf{t}}^s-\frac{{{\bf{f}}_{\bf{t}}^s}}{2\tau_{\rm{t}}}\right\},
\end{equation}
\begin{equation}\label{subrotation}
    \bar{\bf{r}}^{s}=\underset{\left\{{\bf{r}}_m\in\mathcal{R}\right\}_{\forall m}}{\rm argmax}\bar{f}^{s}_{\bf r}\left({\bf{r}}\right)=\mathcal{B}_{\rm{r}}\left\{{\bf{r}}^s-\frac{{{\bf{f}}_{\bf{r}}^s}}{2\tau_{\rm{r}}}\right\},
\end{equation}
\begin{equation}\label{sublongtimepara}
    \bar{\bf{c}}^{s}=\underset{\bf c}{\rm argmax}\bar{f}^{s}_{\bf c}\left({\bf{c}}\right)={\bf{c}}^s-\frac{{{\bf{f}}_{\bf{c}}^s}}{2\tau_{\rm{c}}}.
\end{equation}
$\mathcal{B}_{\rm{t}}\{{\bf{t}}\}$ is a boundary projection function which ensures each element of the vectors ${\bf{t}}$ being projected into its feasible region, which is given by
\begin{equation}
    \left[\mathcal{B}_{\rm{t}}\left\{{\bf{t}}\right\}\right]_i=\left\{\begin{aligned}
        &\left[{\bf{t}}\right]_i^{\rm{L}},~{\rm{if}}~\left[{\bf{t}}\right]_i\leq\left[{\bf{t}}\right]_i^{\rm{L}},\\
        &\left[{\bf{t}}\right]_i,~{\rm{if}}~\left[{\bf{t}}\right]_i^{\rm{L}}<\left[{\bf{t}}\right]_i\leq\left[{\bf{t}}\right]_i^{\rm{U}},\\
        &\left[{\bf{t}}\right]_i^{\rm{U}},~{\rm{if}}~\left[{\bf{t}}\right]^{\rm{U}}_i<\left[{\bf{t}}\right]_i,
    \end{aligned}\right.
\end{equation}
where $[{\bf t}]_i^{\rm L}$ and $[{\bf t}]_i^{\rm U}$ denote the lower and upper bounds on the $i$-th element of ${\bf t}$, respectively, as specified by the movable region $\{\{{\mathcal{T}_{m,n}}\}_{n=1}^{N}\}_{m=1}^M$. The boundary function of the array orientations, $\mathcal{B}_{\rm r}\{{\bf r}\}$, is defined in a similar manner, with the rotatable region predetermined by $\mathcal{R}$.

Finally, given $\bar{\bf{t}}^s$, $\bar{\bf{r}}^s$, and $\bar{\bf{c}}^s$, the antenna positions, array orientations, and long-timescale parameter are updated in the next iteration as
\begin{equation}\label{updatet}
    {\bf{t}}^{s+1}=\left(1-\gamma^s\right){\bf{t}}^s+\gamma^s\bar{\bf{t}}^s,
\end{equation}
\begin{equation}\label{updater}
    {\bf{r}}^{s+1}=\left(1-\gamma^s\right){\bf{r}}^s+\gamma^s\bar{\bf{r}}^s,
\end{equation}
\begin{equation}\label{updatec}
    {\bf{c}}^{s+1}=\left(1-\gamma^s\right){\bf{c}}^s+\gamma^s\bar{\bf{c}}^s,
\end{equation}
respectively, where the sequence $\gamma^s$ follows the same properties of $\rho^s$ and satisfies $\lim\limits_{s\rightarrow\infty}\frac{\gamma^s}{\rho^s}=0$ \cite{LAcssca}. 

\begin{remark}
    {As iteration number $s$ increases, the term ${\bf f}_{\bf v}^s$ in \eqref{gradient} approximates the first-order gradient of the objective function in \eqref{longtermparaformulate} with increasing accuracy. Notably, the convex surrogate functions are constructed as first-order Taylor expansions of the original non-concave functions, augmented by a quadratic regularization term to ensure uniform strong convexity. Based on this construction of surrogate functions, the updates of $\{{\bf t}^s, {\bf r}^s, {\bf c}^s\}$ can be interpreted as the estimation of stationary solutions for the original problem in \eqref{longtermparaformulate}.}
\end{remark}

\begin{algorithm}[t!]
    \renewcommand{\algorithmicrequire}{\bf{Input:}}
    \renewcommand{\algorithmicensure}{\bf{Output:}}
    \caption{CSSCA-based Two-timescale Design}
    \label{CSSCA}
    \begin{algorithmic}[1]
    \REQUIRE${\bf{t}}^0,{\bf{r}}^0,{\bf c}^0$.
    \ENSURE ${{\bf{t}}},{\bf{r}},{\bf c}$.
    \STATE Set iteration index $s=0$.
    \REPEAT
    \STATE Generate a sample of random path coefficients ${\bm{\Psi}}^s$.
    \STATE Obtain the short-timescale receive beamformers ${\bf{w}}_{k,m}\{{\bf{H}}({\bf{t}}^s,{\bf{r}}^s,{\bm{\Psi}}^s),{\bf c}^s\}$ according to \eqref{closeformreceive beamformer}.
    \STATE Construct the surrogate function $\bar{f}^{s}\left({\bf{t}},{\bf{r}},{\bf{c}},{\bm{\Psi}}^s\right)$ according to \eqref{surrogate}.
    \STATE Obtain the stationary points $\bar{\bf t}^s$, $\bar{\bf r}^s$, and $\bar{\bf c}^s$ according to \eqref{subposition}, \eqref{subrotation}, and \eqref{sublongtimepara}.
    \STATE Update the long-timescale variables ${\bf t}^{s+1}$, ${\bf r}^{s+1}$, and ${\bf c}^{s+1}$ according to \eqref{updatet}, \eqref{updater}, and \eqref{updatec}.
    \STATE Set $s=s+1$.
    \UNTIL the maximum iteration number $s_{\rm{max}}$ is reached.
    \end{algorithmic}
\end{algorithm}

The overall two-timescale workflow is summarized in Algorithm~\ref{CSSCA}. The convergence is guaranteed since the iteratively updated solutions converge almost surely to stationary points of the original problem~\eqref{longtermparaformulate} as the iteration index $s \rightarrow \infty$. A detailed proof can be found in Theorem $1$ of~\cite{LAcssca}. The computational complexity of Algorithm \ref{CSSCA} is primarily attributed to the calculation of the gradients $\nabla_{\bf{v}}g\left({\bf t}^s,{\bf r}^s,{\bf c}^s,{\bm \Psi}^s\right),{\bf v}\in\{{\bf t},{\bf r},{\bf c}\}$. The corresponding computational orders are $\mathcal{O}(MN^2(\sum_{k}\sum_mL_{k,m}+M(N^2+KN+K^2))$, $\mathcal{O}(N^2(\sum_{k}\sum_mL_{k,m}+M(N^2+KN+K^2))$, and $\mathcal{O}(K^3MN)$, respectively. Considering the dominant operations, the overall computational complexity of Algorithm~\ref{CSSCA} is given by $\mathcal{O}(s_{\rm{max}}(MN^2(\sum_{k}\sum_mL_{k,m}+M(N^2+KN+K^2)))$. It is worth noting that this long-timescale optimization is executed only when the channel statistical characteristics vary significantly. As a result, the proposed framework effectively reduces the implementation complexity and overhead, in contrast to conventional designs that rely on instantaneous \ac{csi} and require frequent update of antenna positions and array orientations.
\section{Numerical Results}
\subsection{Simulation Setup}
\begin{table}[ht]
\centering
\caption{Simulation parameters}
\label{table:1}
\begin{tabular}{|c|c|}
\hline
{\bf{Parameter}} & {\bf{Value}}  \\ \hline
Carrier frequency & $f_c=20$ GHz\\ \hline
Wavelength & $\lambda=0.015$ m\\ \hline
Number of APs & $M = 10$ \\ \hline
Number of antennas per AP & $N = 6$  \\ \hline
Number of users & $K = 10$  \\ \hline
Number of paths & $L = 6$ \\ \hline
Rician factor & $\kappa=10$ dB\\ \hline
Movable region size for each antenna & $\lambda\times \lambda$\\ \hline
Rotatable range for each array & $[-30^\circ,30^\circ]$\\ \hline
Average noise power & $-70$ dBm\\\hline
Searching step size & $\hat\kappa_{\bf{c}}=\hat\kappa_{\bf{r}}=10$\\\hline
Error tolerance & $\varepsilon=10^{-3}$\\\hline
Maximum number of iterations & $s_{\rm{max}}=100$\\\hline
\end{tabular}
\end{table}

 In this section, numerical results are provided to evaluate the performance of the proposed two-timescale decentralized design for the \ac{6dma}-aided cell-free \ac{mimo} system. We consider the \acp{ap} are deployed uniformly along a ring of radius $140$ m to serve the users distributed within a circular area of radius $120$ m. The path-loss is computed based on the 3GPP Urban Microcell model \cite{3GPPsimu}, given by
\begin{equation*}
    \rho_{k,m}=-22.7-36.7\lg\left(\frac{d_{k,m}}{1~{\rm{m}}}\right)-26\lg\left(\frac{f_c}{1~\rm{GHz}}\right)~[\rm{dB}],
\end{equation*}
where $d_{k,m}$ denotes the distance between the $k$-th user and the $m$-th \ac{ap}, with a height difference of $10$ m. The numbers of received paths between all \acp{ap} and users are assumed to be identical, i.e., $L_{k,m}=L$. The fading coefficients follow $\psi_{k,1,m}\sim\mathcal{CN}(0,(\kappa\rho_{k,m})/({\kappa+1}))$ and $\psi_{k,l,m}\sim\mathcal{CN}(0,\rho_{k,m}/(({\kappa+1})(L-1)))$ for $\forall l\neq1$, where the Rician factor $\kappa$ characterizes the relative strength of the \ac{los} component. The AoAs of the non-LoS (NLoS) paths for each \ac{ap} are generated according to a joint probability density function $f(\theta_{k,l,m},\phi_{k,l,m})=\frac{\sin\theta_{k,l,m}}{2\pi}$ with $\theta_{k,l,m}\in[0,\pi]$ and $\phi_{k,l,m}\in[0,\pi]$ \cite{lpzhuMA}. For each \ac{ap}, the initial spacing between adjacent antennas is set to $5\lambda$. All antennas share the same movable region relative to their initial positions, and all arrays have the same allowable pitch, roll, and yaw rotation ranges. The performance is evaluated through $200$ independent realizations of user distributions and channel coefficients. For each realization, the ergodic sum rate of all users is obtained using $s_{\rm{max}} = 100$ stochastic samples to execute Algorithm \ref{CSSCA}. In the short timescale, the expectations of the stochastic parameters are approximated using the \ac{saa} method. The parameters for \ac{cssca} algorithm are set to $\rho^s=\frac{1}{(1+s)^{0.9}}$ and $\gamma^s=\frac{15}{15+s}$ \cite{LAcssca}.

\begin{figure}[!t]
    \centering
    \subfigure[]{
        \includegraphics[width=0.46\linewidth]{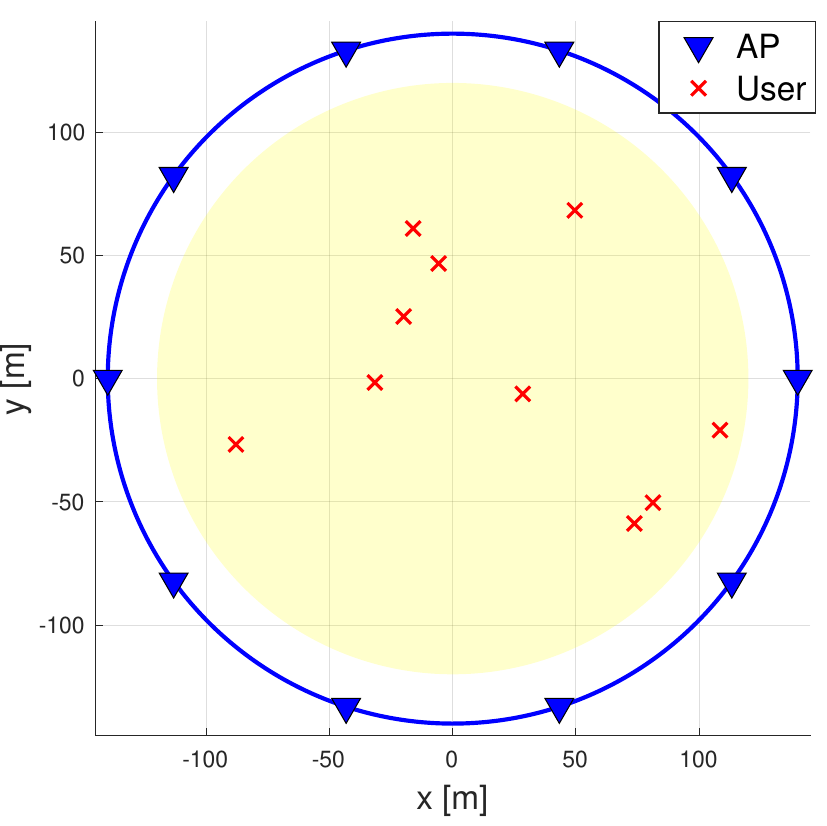}
        \label{uulocation}
    }
    \subfigure[]{
        \includegraphics[width=0.46\linewidth]{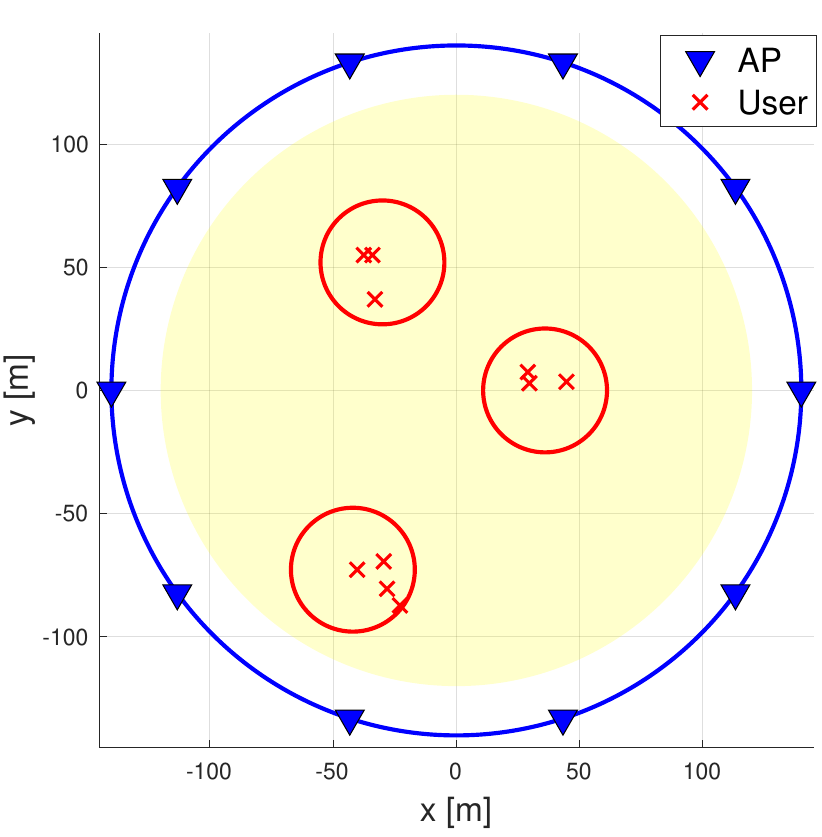}
        \label{dulocation}
    }
    \caption{The simulation setups under different user distributions. (a) Users are uniformly distributed within serving area. (b) Users are densely concentrated around three hotspots.}
    \label{user_distribution}
\end{figure}

\subsection{User Distribution and Baseline Schemes}
To investigate the effectiveness of antenna movement for interference mitigation and performance enhancement under different levels of inter-user interference, particularly in the presence of dominant \ac{los} components, the proposed \ac{cssca} algorithm (labeled as {\bf{Proposed LMMSE with 6DMA}}) is evaluated in two specific communication scenarios, as illustrated in Fig. \ref{user_distribution}, where users are either uniformly distributed within the serving area or densely concentrated around three hotspots. To verify the effectiveness of the proposed \ac{6dma}-aided cell-free system, the performance of Algorithm \ref{CSSCA} is compared with the following baselines: 
\begin{itemize}
    \item {\bf{Centralized MMSE with FPA}}: All \acp{ap} are equipped with \acp{fpa} and upload their acquired \ac{csi} to the \ac{cpu}, which jointly decodes the received signals based on the \ac{mmse} receive beamformer ${\bf W}=({\bf{H}}{\bf H}^{\rm H}+\sigma^2{\bf I}_{MN})^{-1}{\bf H}$ \cite{SQJWMMSE}. The \acp{fpa} within each \ac{ap} are arranged with an inter-element spacing of $5\lambda$, which is the same as the initial position of the \ac{6dma}.
    \item {\textbf{\ac{lmmse} with FPA}}: All \acp{ap} are equipped with \acp{fpa} and independently computes its \ac{lmmse} receive beamformer. The long-timescale parameter ${\bf c}$ is obtained by \ac{saa} method over the long timescale. 
    \item {\bf{LMMSE with flexible position}}: The decentralized cell-free system executes Algorithm~\ref{CSSCA} while keeping the array orientations of all \acp{ap} fixed.
    \item {\bf{LMMSE with flexible orientation}}: The decentralized cell-free system executes Algorithm~\ref{CSSCA} while keeping the antenna positions of all \acp{ap} fixed.
\end{itemize}

\begin{figure}[!t]
    \centering
    \includegraphics[width=0.55\linewidth]{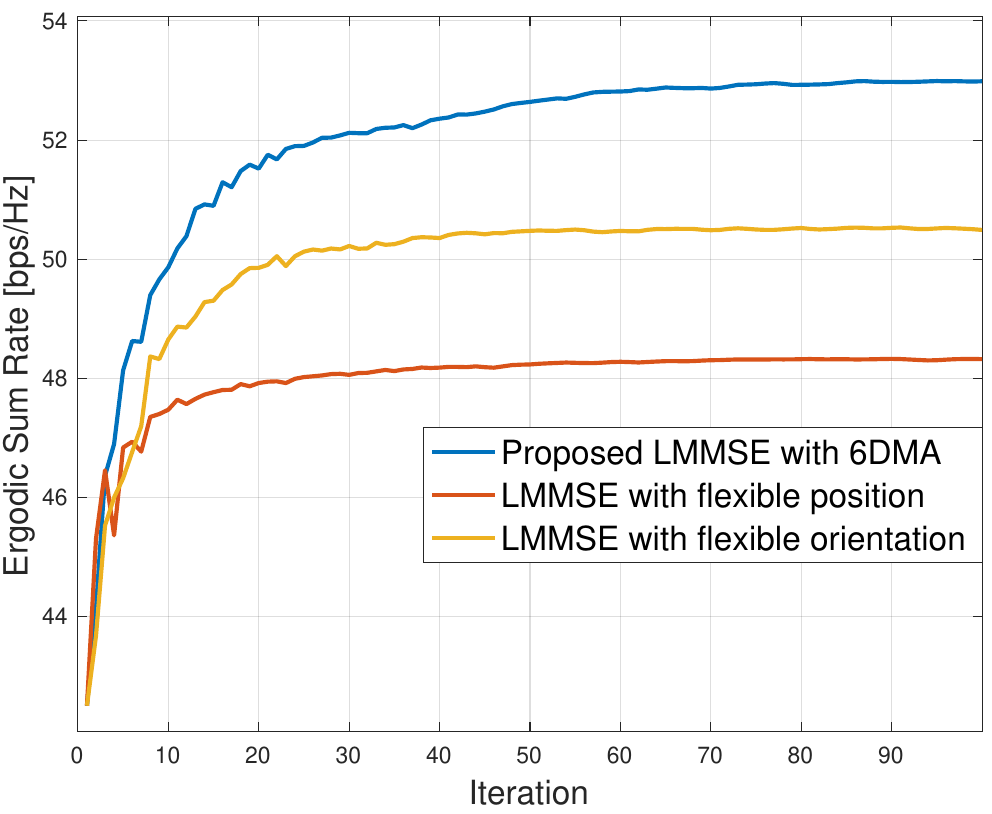}
    \caption{Convergence of Algorithm \ref{CSSCA} under uniform user distribution.}
    \label{convergence}
\end{figure}

\subsection{Algorithm Convergence}
Fig. \ref{convergence} illustrates the convergence behavior of the proposed \ac{cssca}-based algorithm under different flexibilities of antenna movement and array rotation. As observed, although the performance curves exhibit fluctuations, the ergodic sum rate almost surely converges to a steady value as the iteration number increases, which verifies the feasibility of Algorithm \ref{CSSCA}. Specifically, all three benchmarks achieve most of their performance gain within $50$ iterations, indicating that the proposed algorithm features fast convergence and high computational efficiency.

\begin{figure*}[!t]
    \centering
    \subfigure[]{
        \includegraphics[width=0.23\linewidth]{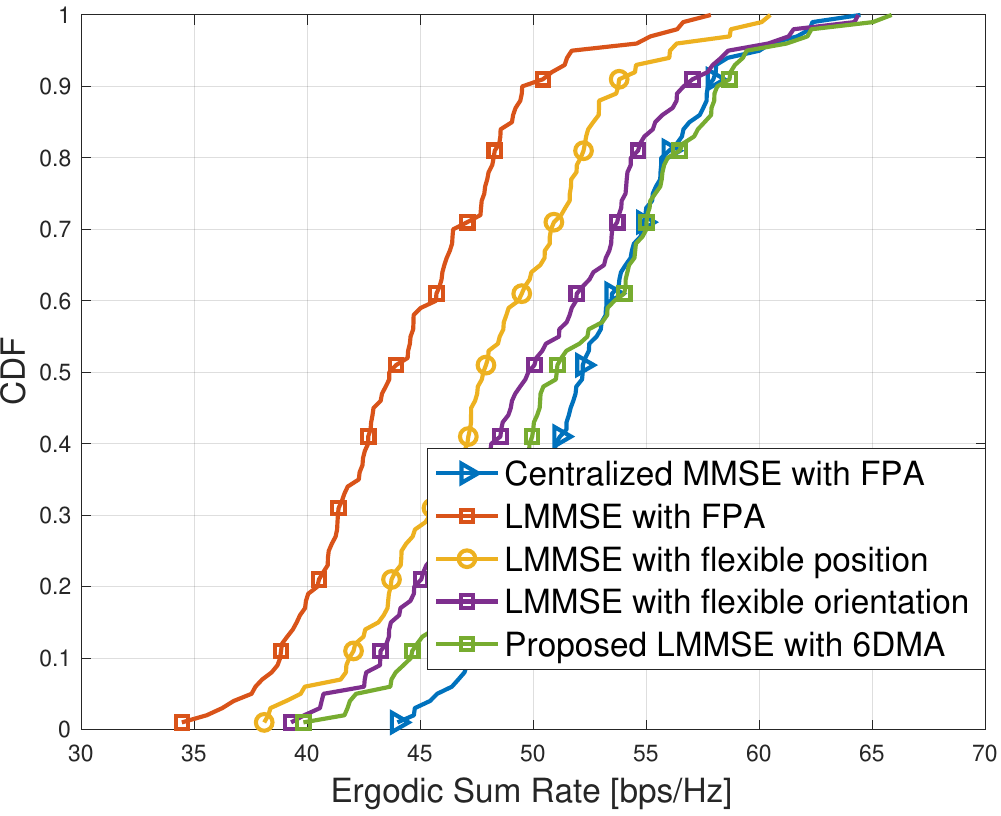}
        \label{uniformcdf}
    }
    \subfigure[]{
        \includegraphics[width=0.23\linewidth]{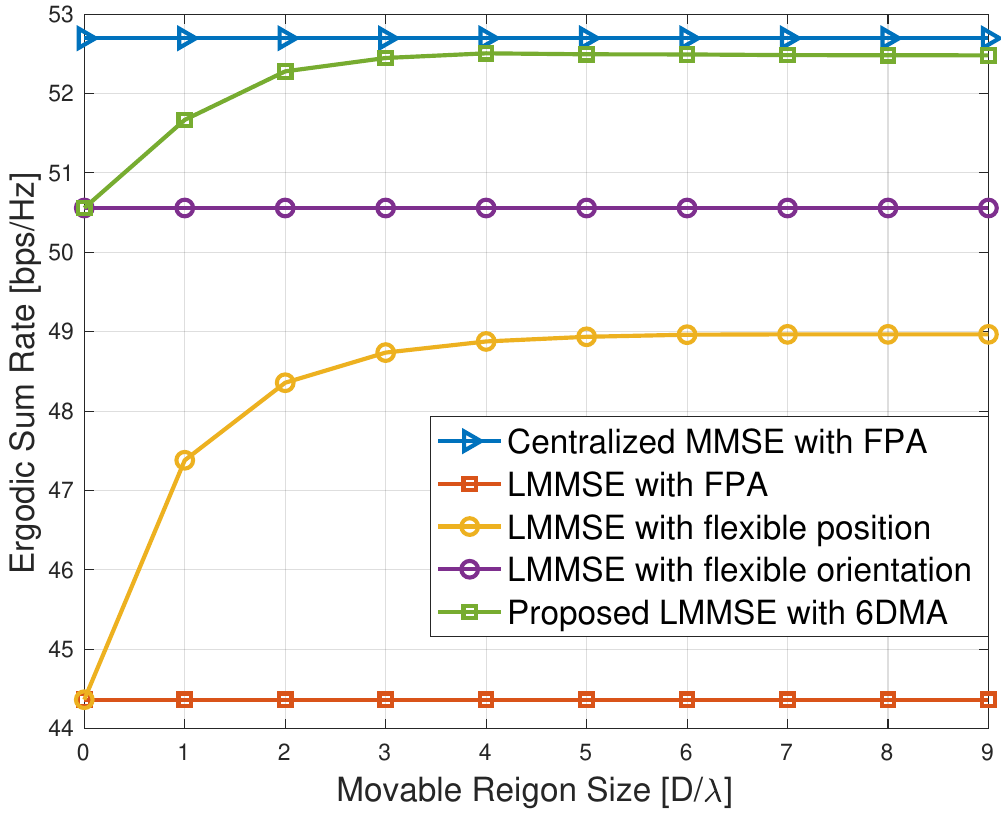}
        \label{unimovable}
    }
    \subfigure[]{
        \includegraphics[width=0.23\linewidth]{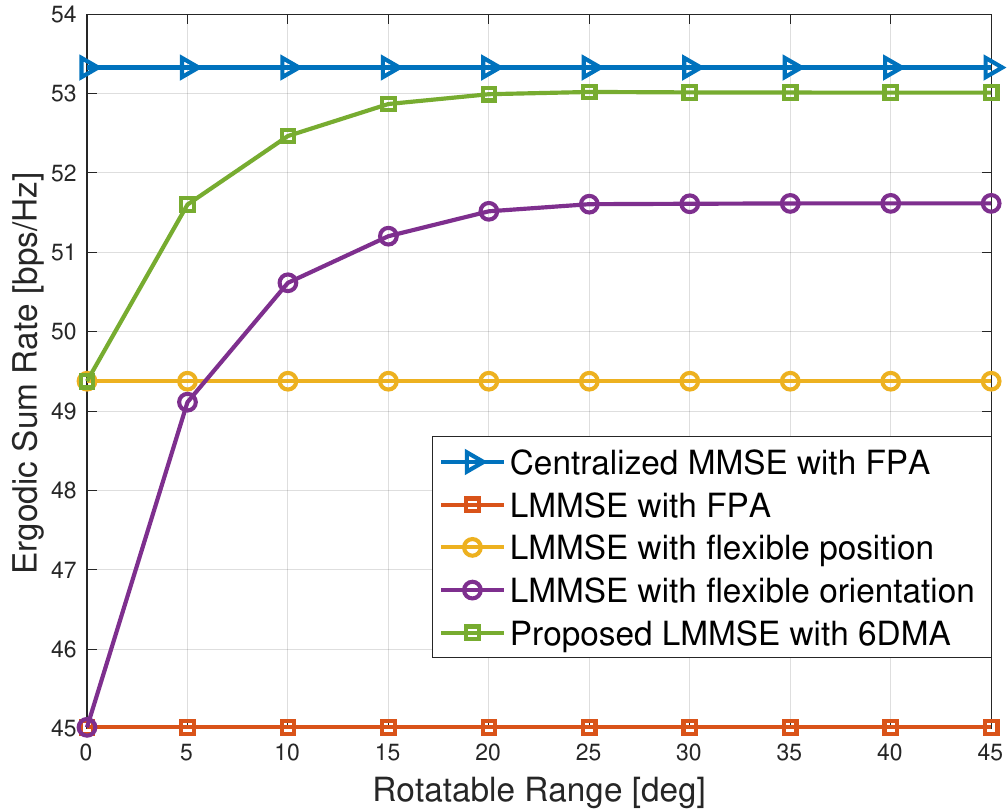}
        \label{unirotatable}
    }
    \subfigure[]{
        \includegraphics[width=0.23\linewidth]{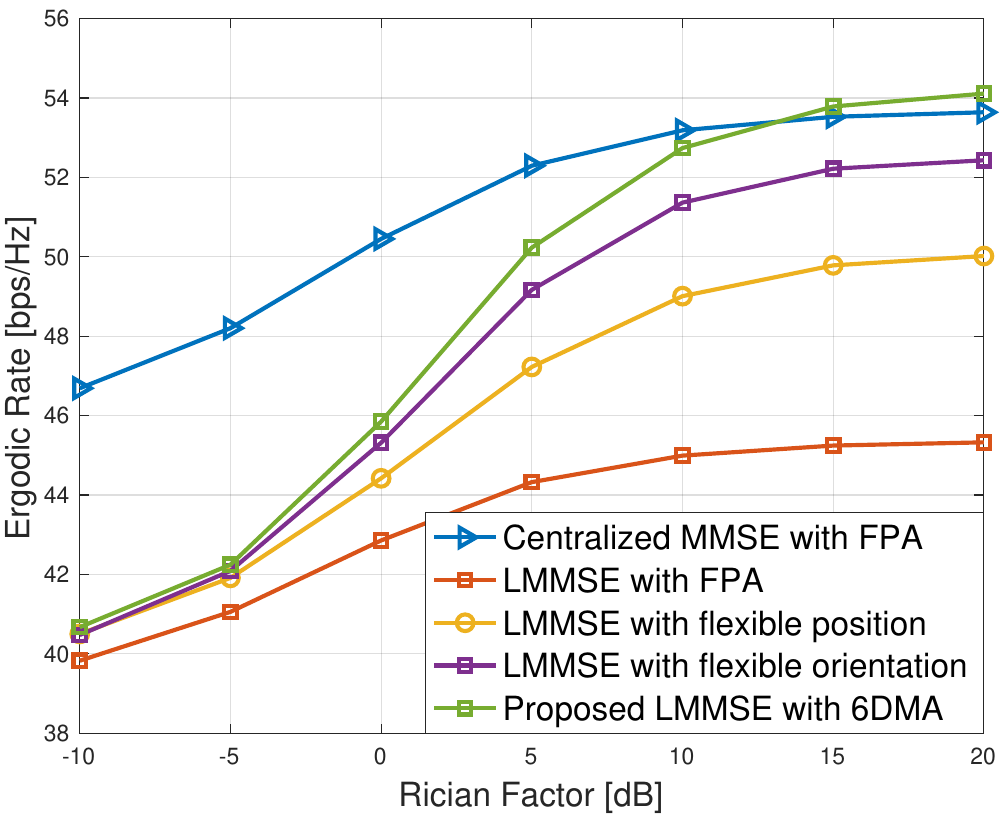}
        \label{unirician}
    }
    \caption{The performance comparison among different schemes with users uniformly distributed within the serving area. (a) The CDF of the ergodic sum rate of all users. (b) The ergodic sum rate of all users versus the movable region size. (c) The ergodic sum rate of all users versus the rotatable range. (d) The ergodic sum rate of all users versus the Rician factor.}
    \label{uniform}
\end{figure*}
\begin{figure*}[!t]
    \centering
    \subfigure[]{
        \includegraphics[width=0.23\linewidth]{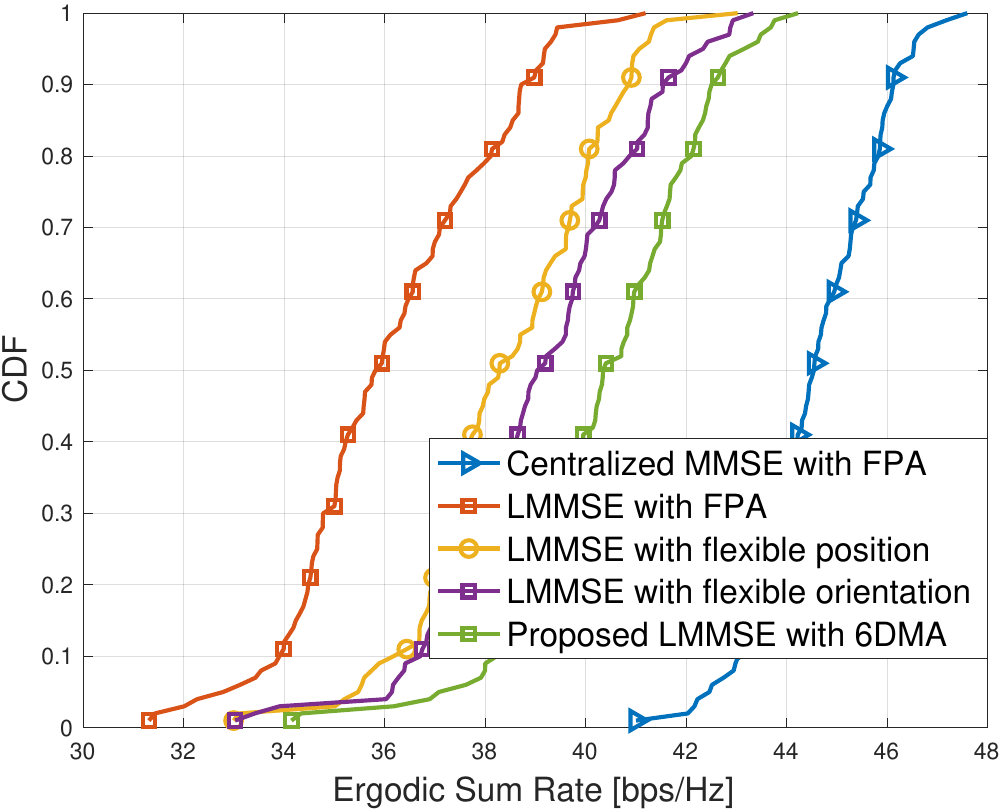}
        \label{densecdf}
    }
    \subfigure[]{
        \includegraphics[width=0.23\linewidth]{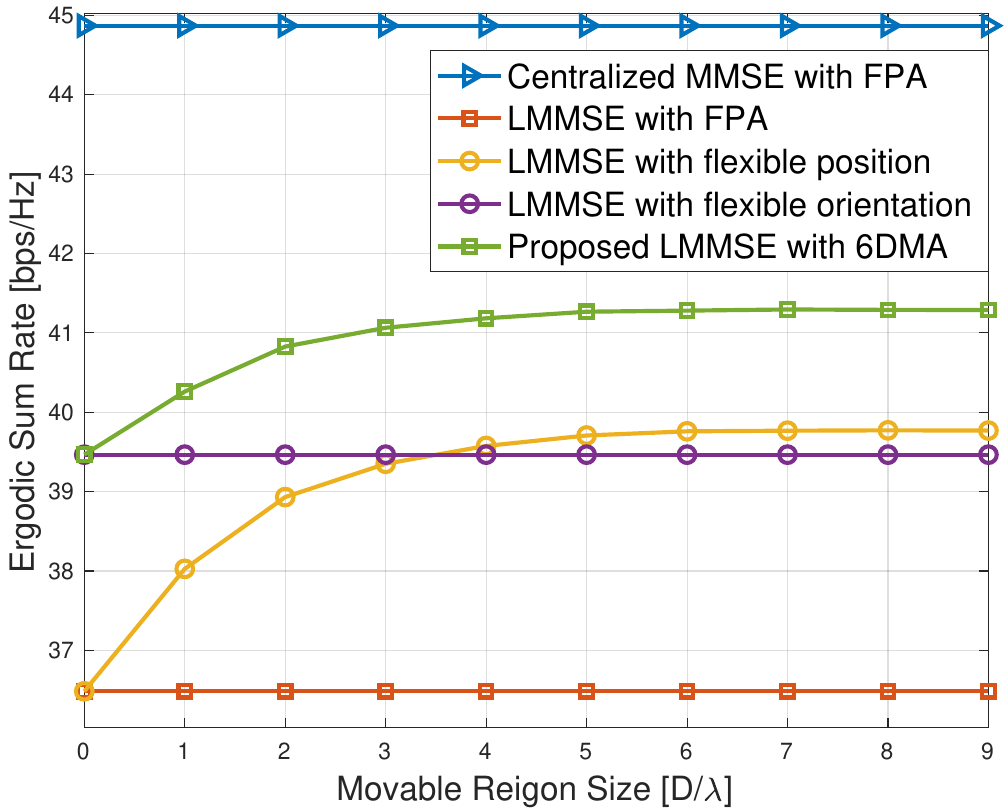}
        \label{densemovable}
    }
    \subfigure[]{
        \includegraphics[width=0.23\linewidth]{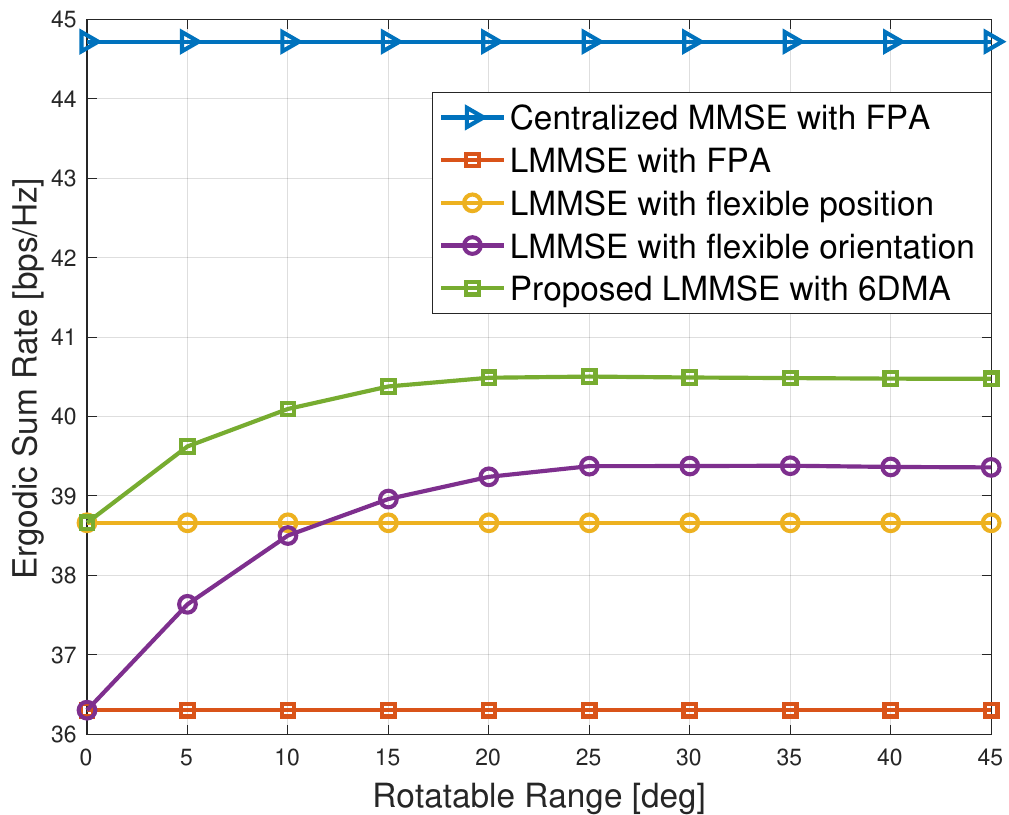}
        \label{denserotatable}
    }
    \subfigure[]{
        \includegraphics[width=0.23\linewidth]{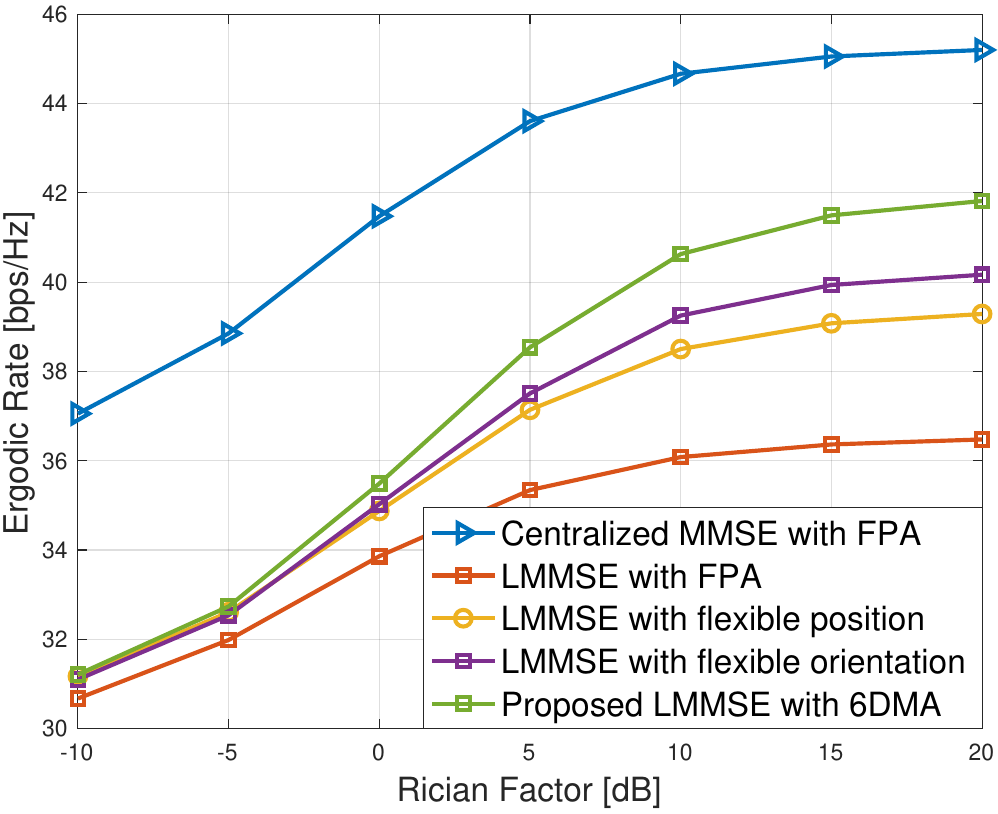}
        \label{denserician}
    }
    \caption{The performance comparison among different schemes with users users concentrated around three hotspots. (a) The CDF of the ergodic sum rate of all users. (b) The ergodic sum rate of all users versus the movable region size. (c) The ergodic sum rate of all users versus the rotatable range. (d) The ergodic sum rate of all users versus the Rician factor.}
    \label{dense}
\end{figure*}

\subsection{Uniform User Distribution}\label{IV-E}
In this subsection, we consider the scenario where the users are uniformly distributed within the serving area, as illustrated in Fig. \ref{uulocation}, and evaluate the performance w.r.t. different system parameters.

Fig. \ref{uniformcdf} presents the empirical \ac{cdf} of the ergodic sum rate obtained from multiple independent realizations of quasi-static AoAs and users locations. It is observed that the proposed \ac{6dma}-aided scheme outperforms the \ac{fpa} and other less flexible counterparts. This is because the array rotation focuses energy toward directions with strong channel paths, thereby exploiting angular \ac{dof} more effectively and improving the overall ergodic sum rate. Interestingly, it shows that under a uniform user distribution with moderate inter-user interference, the proposed scheme approaches the performance of the centralized \ac{mmse} baseline and even surpasses it in specific channel realizations. This result highlights that antenna movement enables cell-free \ac{mimo} systems to better adapt to complex propagation environments and significantly enhance communication performance.

Fig. \ref{unimovable} illustrates the ergodic sum rate of all users versus the movable region size. It is observed that the ergodic sum rate of the \ac{6dma} scheme increases with the region size and eventually converge. These results demonstrate that enlarging the movable region provides additional \ac{dof} for reducing channel correlation among users, thereby efficiently mitigating inter-user interference. Moreover, a bounded per-antenna movement region within three to four wavelengths is sufficient for the cell-free \ac{mimo} decentralized architecture to reap most of the performance gain with a moderate system overhead, and can even approach the performance of the centralized \ac{mmse} baseline.

Fig. \ref{unirotatable} shows the ergodic sum rate of all users versus the rotatable range. The ergodic sum rate achieved by the proposed \ac{6dma} scheme increases with the rotation range until reaching a constant. This suggests that a moderate rotation range is sufficient for each \ac{ap} to orient its array toward the statistically dominant propagation direction and concentrate the radiated energy accordingly. Furthermore, in cell-free \ac{mimo} systems with distributed deployment of \acp{ap}, the wider signal coverage leads to more stronger \ac{los} components, which enables \ac{6dma} to more accurately capture statistically favorable propagation directions, thereby further improving communication performance.

Fig. \ref{unirician} depicts the performance gains achieved by \ac{6dma} under different strength levels of \ac{los} components. Note that small and large Rician factors represent weaker and stronger \ac{los} components, respectively. Specifically, when $\kappa=-10$ dB, the \ac{los} path is weaker than any other NLoS paths, corresponding to \ac{los}-blocked scenarios. Under this condition, the performance gains of \ac{6dma} are severely limited, since there is no clearly dominant propagation direction. In contrast, $\kappa=20$ dB implies that the multi-path channel approximates a pure \ac{los} channel. In this case, the results reveal that the performance gain from \ac{6dma} becomes more pronounced and can even outperform the \ac{mmse} baseline due to the existence of an apparently dominant propagation direction, which allows more effective inter-user interference mitigation and radiated energy concentration.

It should be emphasized that although the centralized scheme achieves a superior performance owing to global \ac{csi} availability and instantaneous joint signal processing, it imposes considerable computational and \ac{csi}-exchange overheads, thereby limiting network scalability. In contrast, the proposed decentralized framework offers a more practical balance between performance and implementation complexity.

\subsection{Concentrated User Distribution}
In this subsection, we consider the scenario where the users are concentrated around three predefined hotspots, as illustrated in Fig. \ref{dulocation}, and evaluate the performance improvement trends w.r.t. different system parameters. 

Fig. \ref{densecdf} shows the CDF of the ergodic sum rate of all users. Under a concentrated user distribution, although the proposed scheme maintains performance superiority over other less flexible benchmarks, its performance gap to the \ac{mmse} baseline becomes substantial. The reason behind is that the channels between the densely distributed users and each \ac{ap} are highly correlated, especially in the presence of dominant \ac{los} components, resulting in severe inter-user interference. Under such conditions, it becomes difficult for each \ac{ap} with a limited number of antennas and \ac{lmmse}-based receive beamformer to completely suppress the interference. 

The relationships between the ergodic sum rate of all users and the movable region size, rotation range, and Rician factor are presented in Figs. \ref{densemovable}, \ref{denserotatable}, and \ref{denserician}, respectively. It can be observed that the proposed \ac{6dma} scheme achieves higher ergodic sum rates compared with the \ac{fpa} and other less flexible counterparts, consistent with the conclusions drawn in Section \ref{IV-E}, thereby verifying the effectiveness of the proposed framework. However, as shown in the three subfigures, the performance improvements brought by \ac{6dma} are significantly reduced compared with those under the uniform user distribution scenario. This is because, in the dense user distribution case, the channels of users located within the same hotspot are highly correlated, which is difficult to decorrelate using the limited number of antennas available at each \ac{ap}. Such limitation restricts the exploitation of additional \ac{dof} provided by antenna movement, thereby limiting further performance enhancement.

In addition, it is noteworthy that, under this dense-user condition, the performance improvements contributed individually by antenna position and array orientation become comparable. This observation can be explained as follows. Antenna position optimization helps mitigate channel correlation by providing higher spatial resolution in the angular domain, enabling better separation of users located in similar directions. On the other hand, antenna orientation optimization primarily enhances directional energy focusing toward the statistically dominant propagation paths. Therefore, jointly considering both antenna position and orientation optimization allows the system to combine their complementary advantages, further improving the overall communication performance of \ac{6dma}-aided cell-free \ac{mimo} systems.

\section{Conclusion}
This paper has investigated a \ac{6dma}-aided cell-free system for uplink multi-user communications. To mitigate the overhead caused by frequent antenna movement and \ac{csi} exchange among distributed \acp{ap}, a two-timescale decentralized optimization framework was developed to effectively maximize the ergodic sum rate of all users. In the short timescale, the optimal \ac{lmmse} receive beamformer design rule was derived, enabling each \ac{ap} to design its receive beamformer based on local instantaneous \ac{csi} and global statistical \ac{csi}. In the long timescale, the \ac{cssca} framework was employed, which iteratively constructs a sequence of surrogate problems to approximate the original stochastic objective function. This approach effectively decouples the optimization variables and enables their updates in parallel, thereby yielding computationally efficient solutions. Numerical results demonstrated that the proposed \ac{6dma}-aided cell-free system significantly outperforms the conventional \ac{fpa} and other antenna movement schemes with less flexibility. Notably, under specific practical conditions, the proposed \ac{6dma} scheme can achieve a performance comparable to and even better than the centralized \ac{mmse} baseline with \acp{fpa}. Moreover, the results highlighted that the performance improvements are more pronounced with sparse user distribution. Beyond the scope of this work, other fronthaul architectures such as unidirectional, tree, or star topologies, which enable enhanced \ac{csi} sharing among \acp{ap}, as well as the corresponding decentralized antenna movement designs, can further exploit the potential of cell-free networks. These directions are worth being explored in future work.

\section*{Appendix\\Derivation of \eqref{longtimevariable}}\label{appendixa}
Substituting the closed-form receive beamformer \eqref{closeformreceive beamformer} into the per-AP optimality condition \eqref{receive beamformeropt} yields
\begin{equation}
    \hat{\bf{G}}_m{\bf{c}}_{k,m}-\hat{\bf{G}}_m\left({\bf{e}}_k-\sum_{i\neq m}\mathbb{E}\left[{\mathbf{H}}_i^{\rm{H}}{\mathbf{{G}}}_i{\bf{c}}_{k,i}\right]\right)=0,~\forall m.
\end{equation}
Since the equality holds for all realizations of the local \ac{csi} $\hat{\bf{G}}_m$, we can obtain that
\begin{equation}\label{longtermeq}
    {\bf{c}}_{k,m}+\sum_{i\neq m}{\mathbb{E}}\left[{\mathbf{H}}_i^{\rm{H}}{\mathbf{G}}_i\right]{\bf{c}}_{k,i}={\bf{e}}_k,~\forall m.
\end{equation}
Stacking \eqref{longtermeq} over $m$ and using the definitions of the block matrices ${\bf{U}}$ and ${\bf{V}}$ gives the linear system
\begin{equation}
    \left({\rm{blkdiag}}\left({\bf{U}}-{\mathbf{V}}\right)+{\bf{U}}^{\rm{T}}{\bf{V}}\right){\bf{c}}_k={\bf{U}}^{\rm{T}}{\bf{e}}_k.
\end{equation}
Therefore, it suffices to show that $({\rm{blkdiag}}({\bf{U}}-{\mathbf{V}})+{\bf{U}}^{\rm{T}}{\bf{V}})$ is nonsingular. According to Woodbury matrix identity, i.e., $({\bf I}+{\bf AB})^{-1}={\bf I}-{\bf A}({\bf I}+{\bf BA})^{-1}{\bf B}$, where ${\bf A}$ and ${\bf B}$ are arbitrary matrices, this is equivalent to proving that ${\bf{I}}_{K}+{\bf{V}}({\rm{blkdiag}}({\bf{U}}-{\mathbf{V}}))^{-1}{\bf{U}}^{\rm{T}}$ is invertible. Let ${\bf{V}}_i=\mathbb{E}[{\mathbf{H}}_i^{\rm{H}}{\mathbf{G}}_i]$, $\forall 1\leq i\leq M$, we first prove that ${\bf{I}}_{K}-{\bf{V}}_{i}$ is invertible for all $i$, which is given by
\begin{equation}\label{finitematrix1}
    \begin{aligned}
        {\bf{I}}_{K}-{\bf{V}}_{i}&=\mathbb{E}[{\bf{I}}_{K}-{\mathbf{H}}_i^{\rm{H}}{\mathbf{G}}_i]\\
        &=\mathbb{E}\left[{\bf{I}}_{K}-{\mathbf{H}}_i^{\rm{H}}\left({\mathbf{H}}_i{\mathbf{H}}_i^{\rm{H}}+\sigma^2{\bf{I}}_N\right)^{-1}{\mathbf{H}}_i\right]\\
        &\stackrel{(b)}{=}\mathbb{E}\left[\left({\bf{I}}_{K}+\frac{1}{\sigma^2}{\mathbf{H}}_i^{\rm{H}}{\mathbf{H}}_i\right)^{-1}\right],
    \end{aligned}
\end{equation}
where equality $(b)$ is obtained by applying the Woodbury matrix identity. Since expectation preserves positive definiteness, it then follows that ${\bf I}_{K}-{\bf V}_{i}$ is positive definite and invertible. Moreover, using the singular value decomposition, it can be shown that ${\bf V}_{i}$ is invertible. Consequently, by exploiting the specific block structures of ${\bf U}$ and ${\bf V}$, we obtain
\begin{equation}\label{inverse}
\begin{aligned}
    &{\bf{I}}_{K}+{\bf{V}}\left({\rm{blkdiag}}\left({\bf{U}}-{\mathbf{V}}\right)\right)^{-1}{\bf{U}}^{\rm{T}}\\&\stackrel{(c)}{=}{\bf{I}}_{K}+\sum_{i=1}^M{\mathbf{V}}_i\left({\bf{I}}_{K}-{\bf{V}}_{i}\right)^{-1}\\
    &\stackrel{(d)}{=}{\bf{I}}_{K}+\sum_{i=1}^M{\mathbf{P}}_i^{\rm{H}}{\bm{\Sigma}}_i\left({\bf{I}}_{K}-{\bm{\Sigma}}_{i}\right)^{-1}{\mathbf{P}}_i,
\end{aligned}
\end{equation}
where $(c)$ follows from $[{\bf A}_1,{\bf A}_2,\cdots,{\bf A}_N]\times[{\bf B}_1,{\bf B}_2,\cdots,{\bf B}_N]^{\rm T}=\sum_{i=1}^{N}{\bf A}_i{\bf B}_i^{\rm T}$ with ${\bf A}_i\in{\mathbb C}^{m\times n}$ and ${\bf B}_i\in{\mathbb C}^{m\times n}$ being arbitrary matrices. The equation $(d)$ is obtained from the unitary decomposition of the Hermitian matrix ${\bf V}_i={\mathbf P}_i^{\rm H}{\bm \Sigma}_i{\mathbf P}_i$, with ${\mathbf P}_i$ being a unitary matrix and ${\bm \Sigma}_i$ a diagonal matrix. Since both ${\bf I}_{K}-{\bf V}_{i}$ and ${\bf V}_{i}$ are positive definite, equation \eqref{inverse} further shows that ${\bf I}_{K}+{\bf V}({\rm blkdiag}({\bf U}-{\bf V}))^{-1}{\bf U}^{\rm T}$ is positive definite and invertible.

Therefore, we have ${\bf{c}}_k=\left({\rm{blkdiag}}\left({\bf{U}}-{\mathbf{V}}\right)+{\bf{U}}^{\rm{T}}{\bf{V}}\right)^{-1}{\bf{U}}^{\rm{T}}{\bf{e}}_k$, which completes the derivation.

\bibliographystyle{IEEEtran}
\bibliography{main}
\end{document}